\documentclass[twocolumn, twocolappendix, floatfix]{aastex631}
\usepackage{amsmath}

\newcommand{\rowname}[1]

\shorttitle{Metal mixing matters}
\shortauthors{Steinwandel et al.}

\graphicspath{{./}{figures/}}

\begin{document}

\title{Pumping Iron: How turbulent metal diffusion impacts multiphase galactic outflows}

\author[0000-0001-8867-5026]{Ulrich P. Steinwandel}
\affiliation{Center for Computational Astrophysics, Flatiron Institute, 162 5th Avenue, New York, NY 10010, USA}
\email{usteinwandel@flatironinstitute.org}

\author[0000-0002-1619-8555]{Douglas Rennehan}
\affiliation{Center for Computational Astrophysics, Flatiron Institute, 162 5th Avenue, New York, NY 10010, USA}

\author[0000-0003-1053-3081]{Matthew E. Orr}
\affiliation{Center for Computational Astrophysics, Flatiron Institute, 162 5th Avenue, New York, NY 10010, USA}

\author[0000-0003-3806-8548]{Drummond B. Fielding}
\affiliation{Center for Computational Astrophysics, Flatiron Institute, 162 5th Avenue, New York, NY 10010, USA}
\affiliation{Department of Astronomy, Cornell University, Ithaca, NY 14853, USA}

\author[0000-0003-2896-3725]{Chang-Goo Kim}
\affiliation{Department of Astrophysical Sciences, Princeton University, 4 Ivy Lane, Princeton, NJ 08544, USA}

\begin{abstract}

Most numerical simulations of galaxy formation and evolution are unable to properly resolve the turbulent cascade at or below the resolution scale and  turbulence models are required to capture the motion of eddies on those unresolved scales.  In this study, we investigate the impact of turbulent metal diffusion models on multiphase outflows originating from dwarf galaxies ($M_{\rm halo} \sim 10^{10} - 10^{11}$ M$_\odot$).  We use our state-of-the-art numerical model for the formation of single stars and non-equilibrium cooling and hydrogen chemistry.
Our simulations are carried out at a mass resolution of $\sim$1 M$_{\odot}$, where the individual supernova explosions are resolved in terms of hot-phase generation and momentum input. We find that mass, energy, and metal loading factors are only weakly affected by the inclusion of a metal diffusion model. The metal enrichment factor at low altitude above the galactic disk is higher by around 20 per cent when the metal diffusion model is included. Specifically, we find more efficient cooling in the cold interstellar medium, as higher amounts of metals are kept in the cold dense phase. The most striking effect of the metal diffusion model is that, without metal diffusion, there is more rapid cooling in the hot phase and a reduced sound speed by a factor of two. Specifically, we find that the hot phase is more metal enriched in the case without metal diffusion leading to more rapid (over) cooling of that phase which is consistent with the higher sound speed we find in the runs with metal diffusion.
\end{abstract}

\keywords{Galactic winds (572) --- Galaxy evolution (594) --- Hydrodynamical simulations (767) --- Stellar feedback (1602) --- Interstellar medium (847)}

\section{Introduction} \label{sec:intro}

Galactic winds are ubiquitous across all galaxies, and are the result of the concerted death throes of stars. The enormous amounts of combined energy deposited by supernova explosions and stellar winds is sufficient to locally quench star formation in the interstellar medium (ISM), while still being able to reach the gas that surrounds galaxies, the circumgalactic medium \citep[CGM; see reviews by][for example]{Somerville2015, Naab2017}. During the tumultuous journey to the outer halo, galactic-scale winds are able to deposit mass, momentum, and metals along their path -- leading to the complex structure observed in the CGM \citep[e.g.,][]{Heckman2017, Tumlinson2011, Tumlinson2017, Werk2016, Zahedy2021}.  In $\lesssim L^*$ galaxies, galactic-scale outflows are thought to be able to prevent further star formation by halting the gas supply from the cooling CGM \citep{Fielding2017_cgm, Pandya2023, Carr2023, Smith2024, Dave2012}. Evidently, understanding the thermodynamic and structural evolution of these outflows is essential to developing a complete picture of galaxy formation \citep[e.g.,][]{Katz2022, Kim2015_outflows, Kim2022, Gutcke2021, Hu2016, Hu2017, Hu2019, Rey2024, Smith2021, Smith2024, Steinwandel2020, Steinwandel2022, SteinwandelLMC, Steinwandel_Goldberg} .

One important consideration in the feedback-regulated star formation model is that the gas cooling rate in both the ISM and the CGM is metallicity-dependent.  Any increase in local gas-phase metal abundance promotes additional cooling due to the availability of new atomic transitions. The local gas-phase metallicity can vary widely from the global gas-phase average, as both the ISM and the CGM are multiphase media \citep[see, e.g.,][and references therein]{Kim2022, Fielding2020}. In particular, the thermodynamic phase space of the ISM is rich and spans many orders of magnitude in density and temperature \citep{Fielding2020, Kim2023}, while the CGM is equally rich, as evidenced by the complex kinematic structure observed in the COS-Halos survey \citep[e.g.,][]{Tumlinson2017, Prochaska2017, Werk2016}. On even smaller scales, hyper-local fluctuations within each medium can lead to complex metallicity structure -- promoting cooling. Observations support a reasonably well mixed (in terms of metallicity) ISM \citep[e.g.,][]{Esteban2018, Esteban2022, Arellano2021}, while the CGM is more complex due to the interactions with galactic outflows launched from the ISM \citep[e.g.,][]{Faucher2023}  The question then arises: what drives local fluctuations in these gaseous media, and how do galactic outflows impact metal distributions?

The ejecta from SNe and stellar winds not only inject metals into their environments but  also inject sufficient energy to drive turbulence: both locally, in the ISM, and also in the CGM in the form of galactic-scale outflows \citep[e.g.,][]{Prasad2015, Prasad2018, Karen2016, Bourne2017, Fielding2017_cgm, Fielding2018, Sokolowska2018, Li2020}, generating vorticity over a huge range of spatial scales.  How that injected energy evolves over time depends directly on the viscosity of the gas, since any source of viscosity present in the gas acts to convert that injected (kinetic) energy into thermal energy, effectively damping the motion \citep[e.g.,][]{Pope2000}. However, the ISM is effectively inviscid \citep[e.g.,][]{Draine2011} and the injected kinetic energy creates a cascade of inertial motion down to the viscous scale -- the turbulent cascade.  The range of scales in this process is set by the largest (energy-containing) scale, $L$ and the smallest (dissipative) scale, $\eta$.  The dimensionless number describing the degree of turbulence in the medium is the Reynolds number, $Re \sim (L / \eta)^{4/3}$. Typically, $Re \sim 10^3$ indicates the transition to turbulence, whereas $Re$ in the ISM has been estimated to be $\sim10^7$ \citep{Elmegreen2004}. 

The real physical problem arises because the numerical resolution scale $h$ is above the physical dissipation scale $\eta$, usually with $L > h > \eta$.  When this is the case, the turbulent cascade is truncated at scale $\sim h$, causing  kinetic energy to build up to a large scale, $\sim 10h$ \citep[][]{Schmidt2015, Rennehan2019, Rennehan2021}.  The simplest models that ameliorate the issue assume that the small scales are universal and isotropic, and that the motion of the unresolved eddies behaves as a diffusive process, transporting momentum, thermal energy, and metals.  The idea is to add additional terms (i.e., sub-grid models) to the hydrodynamic conservation equations that capture the diffusive properties of the eddies near the grid scale, in an attempt to improve the simulation accuracy on the largest scales\footnote{This is the large-eddy simulation (LES) approach.} \citep[e.g.,][]{Pope2000}. 

The earliest model to improve mixing accuracy came from \citet{Smagorinsky1963}, who proposed a simple model based on eddies moving over a length scale $h$ with a velocity based on the trace-free symmetric shear tensor $|S^*|h$.  The model assumes that shear fluctuations in the gas drive turbulent motions, although there are other methods to determine the degree of turbulence of the gas \citep{Piomelli1994, Rennehan2019, Hu2020, Hopkins2021_extrinsic, Hopkins2022_spectra, Hopkins_fire3}.  

In galaxy formation simulations, the first models were similar to the Smagorinsky model and were motivated by the poor treatment of both boundary-layer mixing and post-shocking mixing in smoothed particle hydrodynamics \citep[SPH;][]{Greif2008, Wadsley2008}.  SPH in particular suffers due to its reliance on constant mass particles: mass flow from unresolved turbulent eddies between particles, including the flow of metals, is impossible \citep{Shen2013}. Metals are of prime importance to galaxy formation, as our observations and theoretical models of the Universe are highly dependent on the metallicity of stars and gas.  Simulations that included these metal diffusion models showed improved accuracy in large-scale mixing \citep{Brook2014, Williamson2016, Su2017, Colbrook2017, Hafen2019, Hafen2020, Rennehan2019, Rennehan2021}, and substantially better matches to observational constraints in dwarf galaxies \citep[e.g.,][]{Escala2018}.

In this paper, we introduce the Smagorinsky model to highly resolved simulations of dwarf galaxies in order to measure the impact of turbulent metal diffusion on the metal phase-space structure in multiphase galactic outflows.  In Sec.~\ref{sec:methods}, we discuss our simulation setup and our implementation of the metal diffusion model.  In Sec.~\ref{sec:results}, we show the impact of the model on the global properties of our simulated dwarf galaxy and on the phase structure of the galactic outflows.  In Sec.~\ref{sec:discussion}, we discuss the results in the context of contemporary simulations.  In Sec.~\ref{sec:conclusions}, we draw conclusions about the future of metal diffusion models in Lagrangian hydrodynamics simulations of galaxy formation.

\section{Numerical Methods and Simulations} \label{sec:methods}

\subsection{Gravity and hydrodynamics}
All the simulations used in this paper are carried out with the Tree-multi-method code P-Gadget3, based on the codes P-Gadget2 \citep{springel05} and Gizmo \citep{Hopkins2015}, with important updates by \citet{Hu2014} and \citet{Steinwandel2020} for the specific version used in this paper. Gravity is computed using a Tree-code \citep[][]{Barnes1986, springel05}. We use the Riemann-solver-based meshless finite mass (MFM) method \citep{Gaburov2011, Hopkins2015} to solve for the fluid flows, as implemented in \citet{Steinwandel2020}. Fluid fluxes are reconstructed with an Harten-Lax-van-Leer contact (HLLC)-Riemann solver between Lagrangian tracer particles. We estimate the surface areas for the flux calculation with the smoothing lengths at the midpoint of individual particles. All fluxes are computed pairwise.

\subsection{Cooling and chemistry network}

The cooling and chemistry network follows the frameworks adopted in the simulations of \citet{Walch2015}, \citet{Girichidis2016}, \citet{Girichidis2018}, \citet{Gatto2017}, \& \citet{Peters2017}, which are themselves based on the works of \citet{Nelson1997}, \citet{Glover2007a}, \citet{Glover2007b} and \citet{Glover2012}. Relevant rate equations have been updated following \citet{Gong2016}, making the framework relatively similar to the recent TIGRESS-NCR framework \citep{Kim2022}. The code has the ability to trace six individual species in our default chemistry network:
\begin{enumerate}
    \item molecular hydrogen (H$_{2}$)
    \item ionized hydrogen (H$^{+}$)
    \item neutral hydrogen (H)
    \item carbon monoxide (CO)
    \item ionized carbon (C$^{+}$)
    \item ionized oxygen (O$^{+}$)
\end{enumerate}
However, we note that we disabled the CO network for all the simulations in this paper because the metal diffusion model in the current version tampers with the tolerance values of the non-equilibrium chemistry solver. 

For this paper, we decided to circumvent this issue by disabling the CO network, since we formed about 12 M$_{\odot}$ of total mass in our WLM-like simulations and about 30 M$_{\odot}$  in our LMC-like simulations. At our given mass resolution of 4 M$_{\odot}$, this amount of mass can be neglected. We note that we also follow the free electrons and assume that all silicon in the simulation is stored as Si$^{+}$. Molecular hydrogen condenses on dust grains; therefore, we solve the non-equilibrium rate equations directly to obtain $x_{H_{2}}$ and $x_{H^{+}}$ and obtain $x_H$ based on the conservation law:
\begin{equation}
    x_\mathrm{H} = 1 - 2 x_{\mathrm{H}_2} - x_{\mathrm{H}^{+}}.
\end{equation}
We track $n_e$ by calculating:
\begin{equation}
    x_\mathrm{e} = x_{\mathrm{H}^{+}} + x_{\mathrm{C}^{+}} + x_{\mathrm{Si}^{+}}
\end{equation}
Non-equilibrium cooling rates are solved by assuming the local density, temperature and chemical abundance of individual gas cells. 

The dominant cooling processes are fine-structure line cooling of atoms and ions (C$^{+}$, Si$^{+}$ and O), vibration and rotation line cooling due to molecules (H$_{2}$ and CO), Lyman-$\alpha$ cooling of hydrogen, collisional dissociation of H$_{2}$, collisional ionization of hydrogen and recombination of H$^{+}$. 

The primary contributions to the ISM heating rate are photoelectric heating, the presence of polycyclic aromatic hydrocarbonates (PAHs), a constant cosmic ray ionization rate of $10^{-19}$ s$^{-1}$, photodissociation of H$_{2}$, UV-pumping of H$_{2}$ and formation of H$_{2}$. 

High-temperature cooling ($T  > 3 \times 10^{4}$ K) is enabled via  \citet{Wiersma2009} as implemented by \citet{Aumer2013} with contributing  elements H, He, C, N, O, Ne, Mg, Si, S, Ca, Fe, and Zn. The metals in our LMC-diff and WLM-diff simulations use a similar metal diffusion model as that implemented in \citet{Aumer2013}. Specifically, we follow the implementation of \citet{Shen2013} by evolving a scalar quantity based on
\begin{align}
    \left(\frac{dA}{dt}\right)_D = \nabla (D\nabla A), 
\end{align}
with a diffusion coefficient D that is adopted based on the theory by \citet{Smagorinsky1963} and defined via: 
\begin{align}
    D = C |S_{ij}^{\star}| h^2,
    \label{eq:diff}
\end{align}
where $|S_{ij}^{\star}|$ denotes the  trace-free shear tensor. This type of metal diffusion model is quite common today in Lagrangian codes at fixed mass resolution. We adopt an efficiency C of 0.05, following the arguments in \citet{Shen2010} based on turbulence theory. We note that one could follow \citet{Rennehan2019} and directly solve for the metal diffusion rate based on the local turbulent properties in the gas. 

However, as pointed out in the follow-up study in \citet{Rennehan2021}, this is much more expensive than the isotropic diffusion model adopted here and provides no improvement in the metal distributions. Nevertheless, such a model would be interesting to pursue in future work, since our simulations have high enough resolution to meaningfully predict the properties of the ISM turbulence. However, it remains unclear, even at our excellent resolution, whether the turbulent cascade is resolved down to the resolution limit of the simulation.

\subsection{Star formation and stellar feedback}

The star formation and stellar feedback models are nearly identical to those in the previous works of \citet{Hu2016}, \citet{Hu2017}, \citet{Hislop2022}, \citet{Lahen2020a}, \citet{Lahen2020b}, \citet{Lahen2022}, \citet{Steinwandel2022}, and \citet{SteinwandelLMC}. We note that we do not adopt a full IMF-sampling approach as recently introduced in \citet{Lahen2023}. Hence, only a short overview is presented here. 
Individual stars are sampled from the initial mass function (IMF) using the method of \citet[][]{Hu2017} with a Schmidt-type star formation relation and a fixed star formation efficiency of $2\%$ per free-fall time (i.e., $\dot \rho_\star = 0.02 \times \rho_{gas}/t_{\rm ff}$). Further, a Jeans mass threshold is applied to handle unresolved high-density gas. All the star particles with masses higher than 4 M$_{\odot}$ are realized as single stars. These single stars contribute to the energetics of the ISM via photoelectric heating, photoionizing radiation, metal enrichment from asymptotic giant branch (AGB) stellar winds and feedback of core-collapse supernovae (CCSNe), based on the model of \citet{Hu2017} for the SPH version of the code and \citet{Steinwandel2020} for the MFM version of the code. Every CCSN event is individually resolved with a feedback energy of $10^{51}$ erg\footnote{See \citet{Steinwandel_Goldberg} for the effects of variation in the SN energy level on our model.}. We adopt realistic lifetimes of massive stars based on \citet{Georgy2013}, allowing us to self-consistently trace the locations in the ISM where SNe explode. 

\subsection{Initial conditions and simulations}

The initial conditions for this simulation were produced using the method described in \citet[][]{springel05b}, which is typically referred to as ``MakeDiskGal'' or ``MakeGalaxy.'' For input parameters, we chose typical values that are in agreement with observations of the Large Magellanic Cloud (LMC) \citep[e.g.,][based on kinematic arguments]{Erkal2019} as well as values previously adopted in numerical simulations to select an LMC-type galaxy \citep[e.g.,][]{Hopkins2018}. For the exact parameters of LMC, see \citet{SteinwandelLMC}. However, we note that to reduce the computational expense, the system is set up to be at the lower limits of the halo mass and gas fraction ranges. Hence, we adopt a total dark matter halo mass of $2 \times 10^{11}$ M$_{\odot}$ and a stellar background potential of $2 \times 10^{9}$ M$_{\odot}$; we also adopt a gas disk with a total mass of $0.5 \times 10^{9}$ M$_{\odot}$ for which we adopt a uniform mass resolution of 4 M$_{\odot}$, resulting in $\sim 10^8$ gas particles ($512^3$) and a mean particle spacing of $\sim$ 1.0 pc (mean effective spatial resolution) for the gaseous body of the galaxy. For the dark matter halo, we adopt a concentration parameter of $c=9$ modeled as a Hernquist profile. We adopt a disc scale length of 1 kpc and a disc scale height of 250 pc, with an initial metallicity of 0.1 $Z_{\odot}$. The gravitational softening parameter of the simulations is set to 0.5 pc for gas and stars throughout the total run time of the simulation (500 Myrs). In total, we carry out four simulations: LMC-nodiff, LMC-diff, WLM-nodiff and WLM-diff. The WLM type simulations are modelled after the local dwarf irregular galaxy Wolf-Lundmark-Melotte that has become a test bed setup in recent years for dwarf galaxies with resolved ISM \citep[see e.g.;][]{Gutcke2021, Hislop2022, Steinwandel2020, Steinwandel2022}. Due to the high computational cost of running the LMC-type analogues for more than 500 Myr, we focus on the time evolution of the WLM-type models, which we can easily carry out for a total time of 1 Gyr. The exact parameters for WLM can be found in \citet{Hu2017} and \citet{Steinwandel_Goldberg} and are not repeated here.

\section{Results} \label{sec:results}

\begin{figure*}
    \centering
    \includegraphics[width=0.49\textwidth]{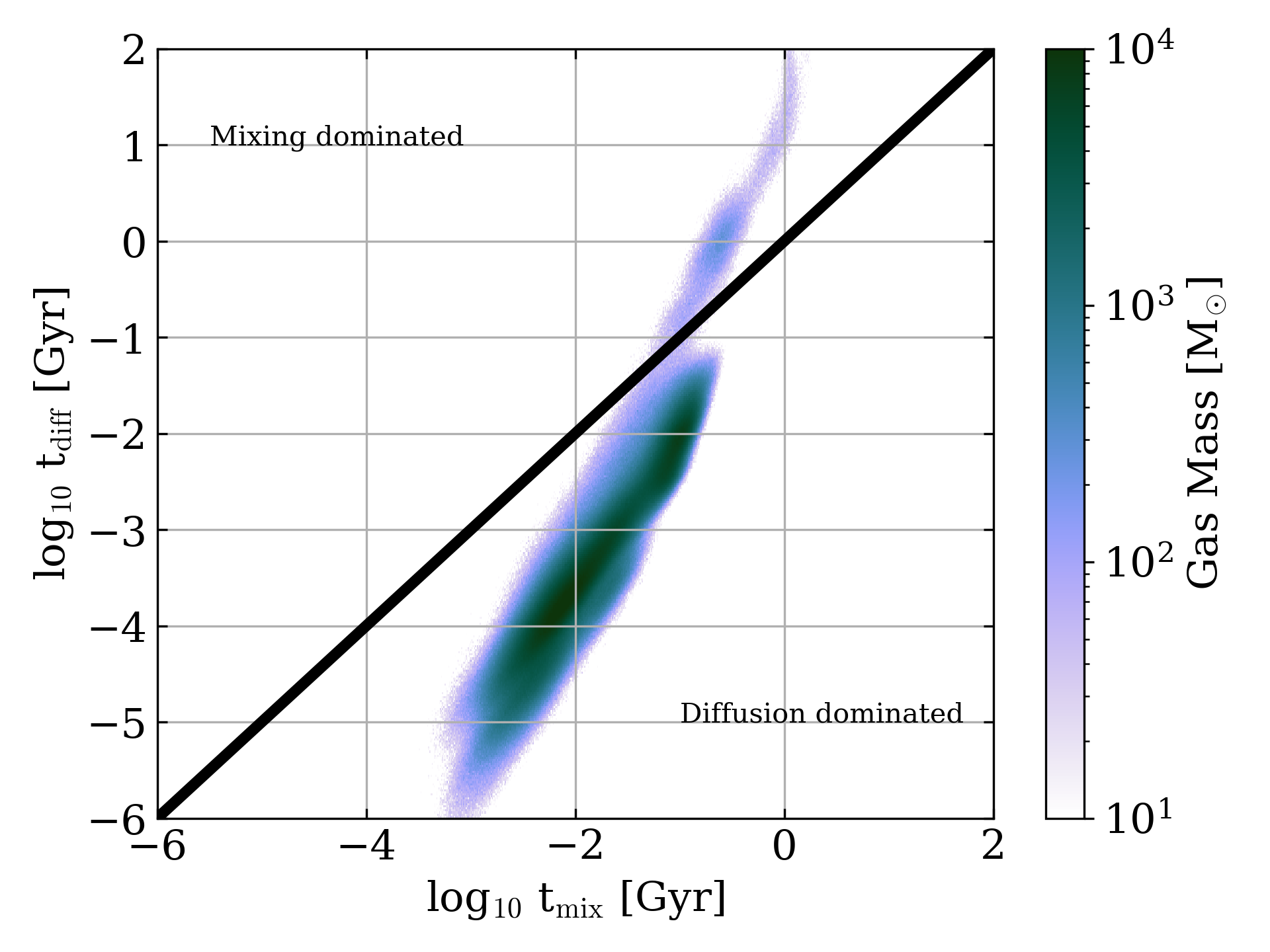}
    \includegraphics[width=0.49\textwidth]{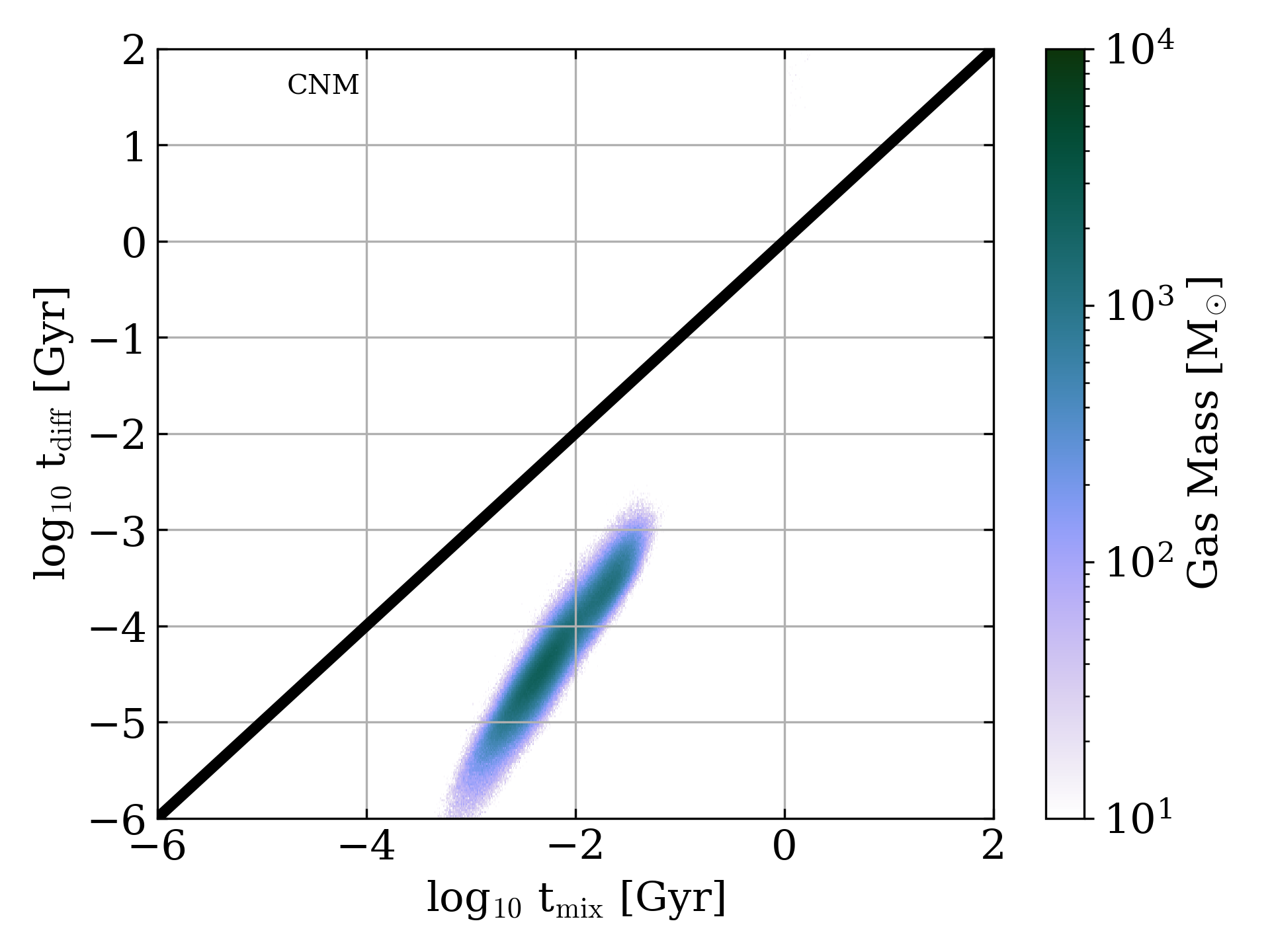 }
    \includegraphics[width=0.49\textwidth]{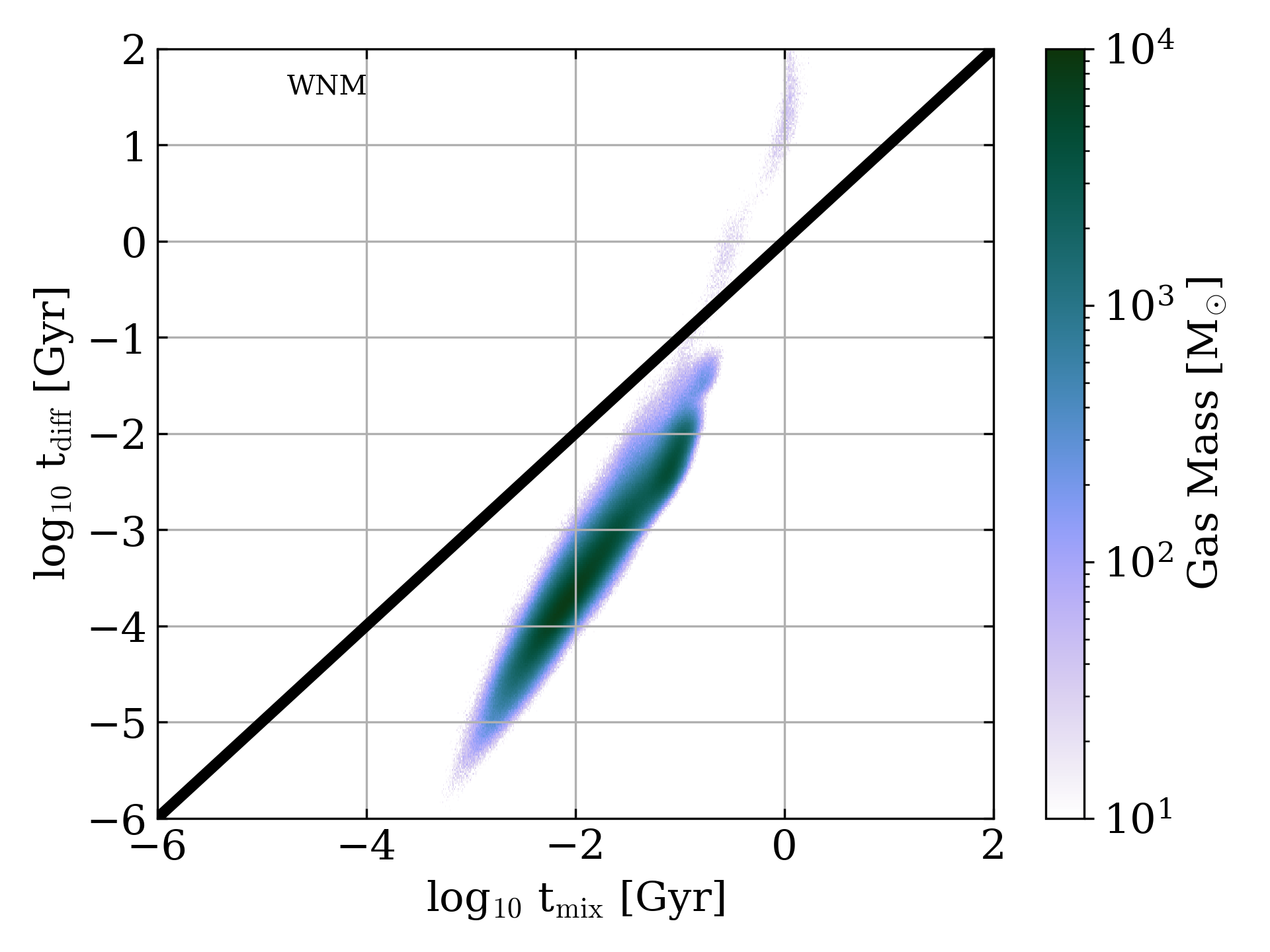 }
    \includegraphics[width=0.49\textwidth]{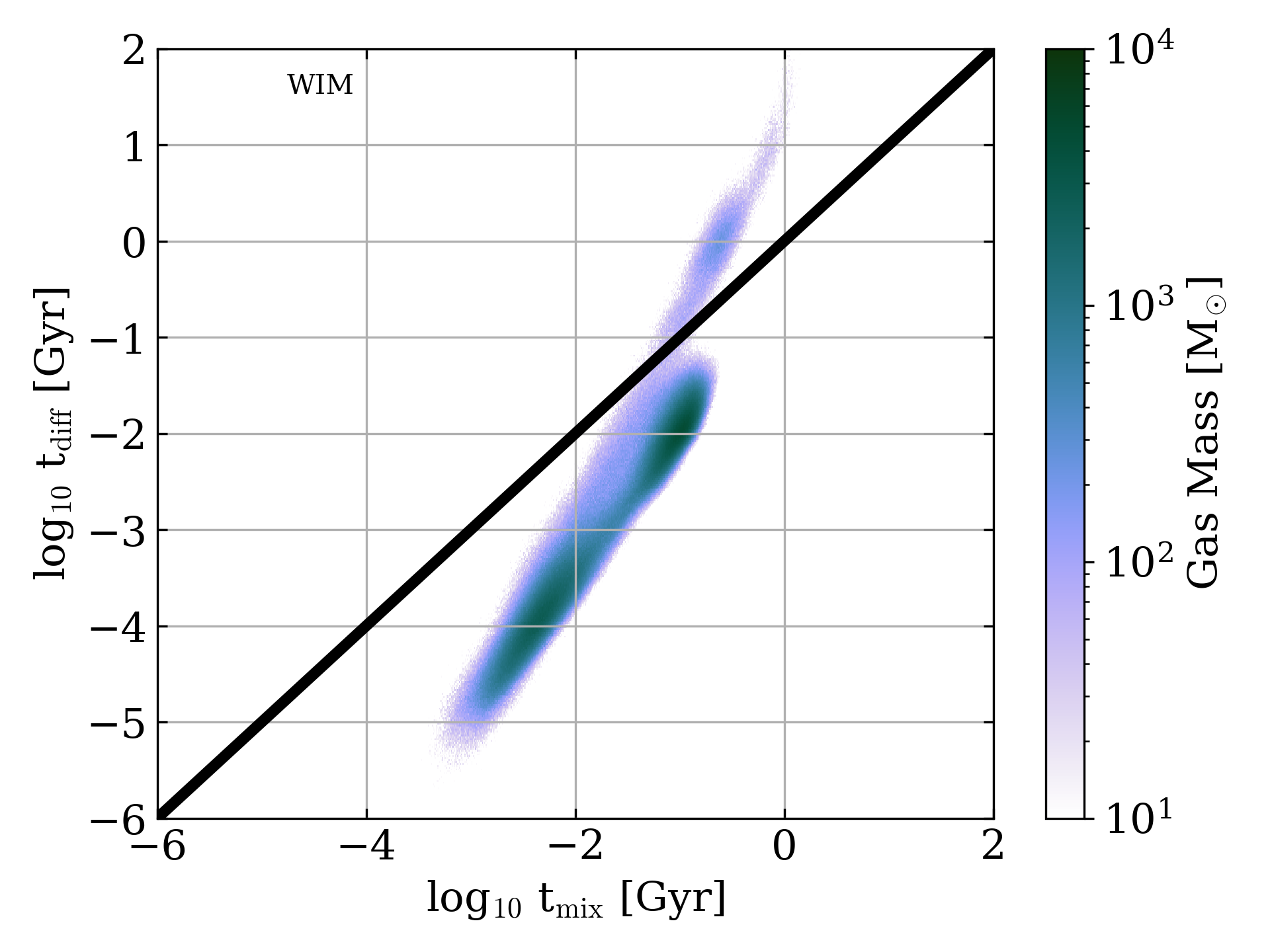 }
    \caption{Mass-weighted joint 2D PDFs of diffusion time vs. turbulent diffusion time for all gas (top left), as well as restricted to the CNM (top right), the WNM (bottom left) and the WIM (bottom right). We note that the HIM carries very little mass and is therefore not displayed in this figure.}
    \label{fig:split_diff}
\end{figure*}

\begin{figure}
    \centering
    \includegraphics[width=0.49\textwidth]{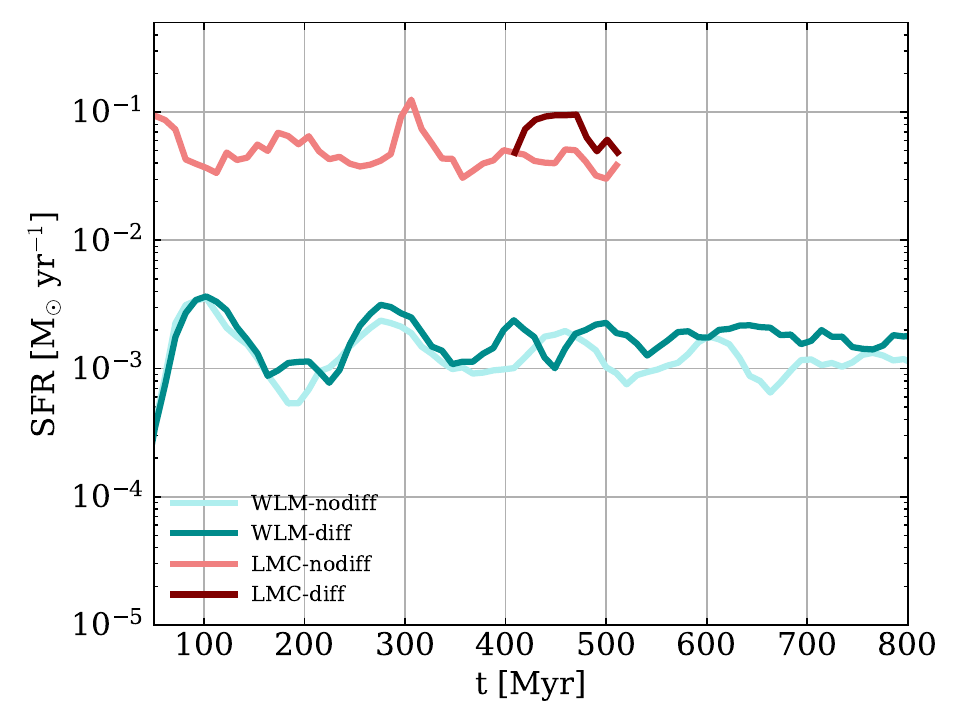}
    \caption{Star formation rate of our four simulated models: WLM-nodiff (light turquoise), WLM-diff (dark turquoise), LMC-nodiff (salmon) and LMC-diff (dark red). We note that we only simulated one wind driving cycle for the model LMC-diff,  starting from a snapshot at 400 Myr of evolution of the run LMC-nodiff and finishing at 500 Myr, and hence the star formation histories of LMC-nodiff and LMC-diff are identical. We did this due to the high computational expense of the runs LMC-nodiff and LMC-diff.}
    \label{fig:sfr}
\end{figure}

\begin{figure*}
    \centering
    \includegraphics[width=0.49\textwidth]{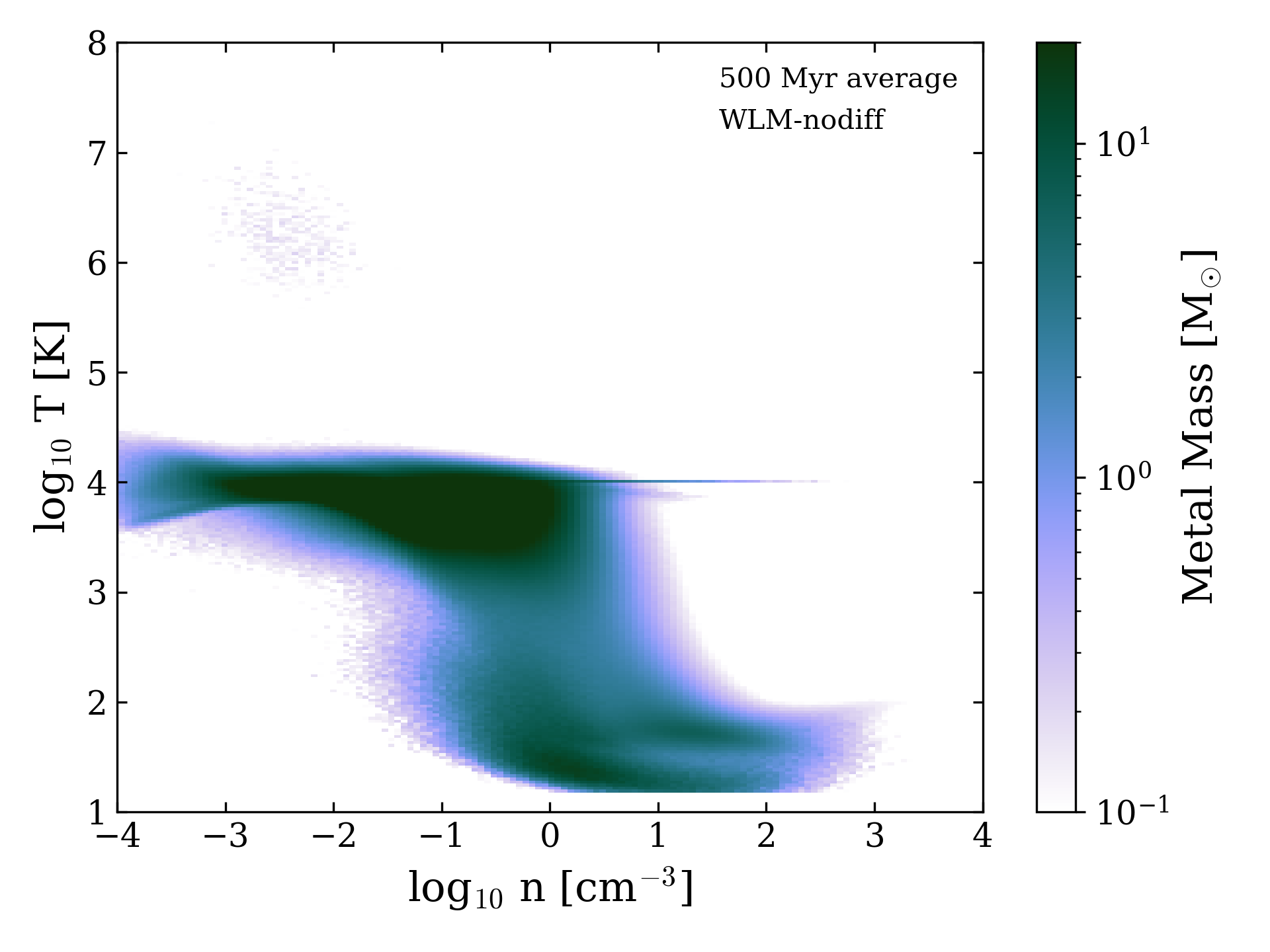}
    \includegraphics[width=0.49\textwidth]{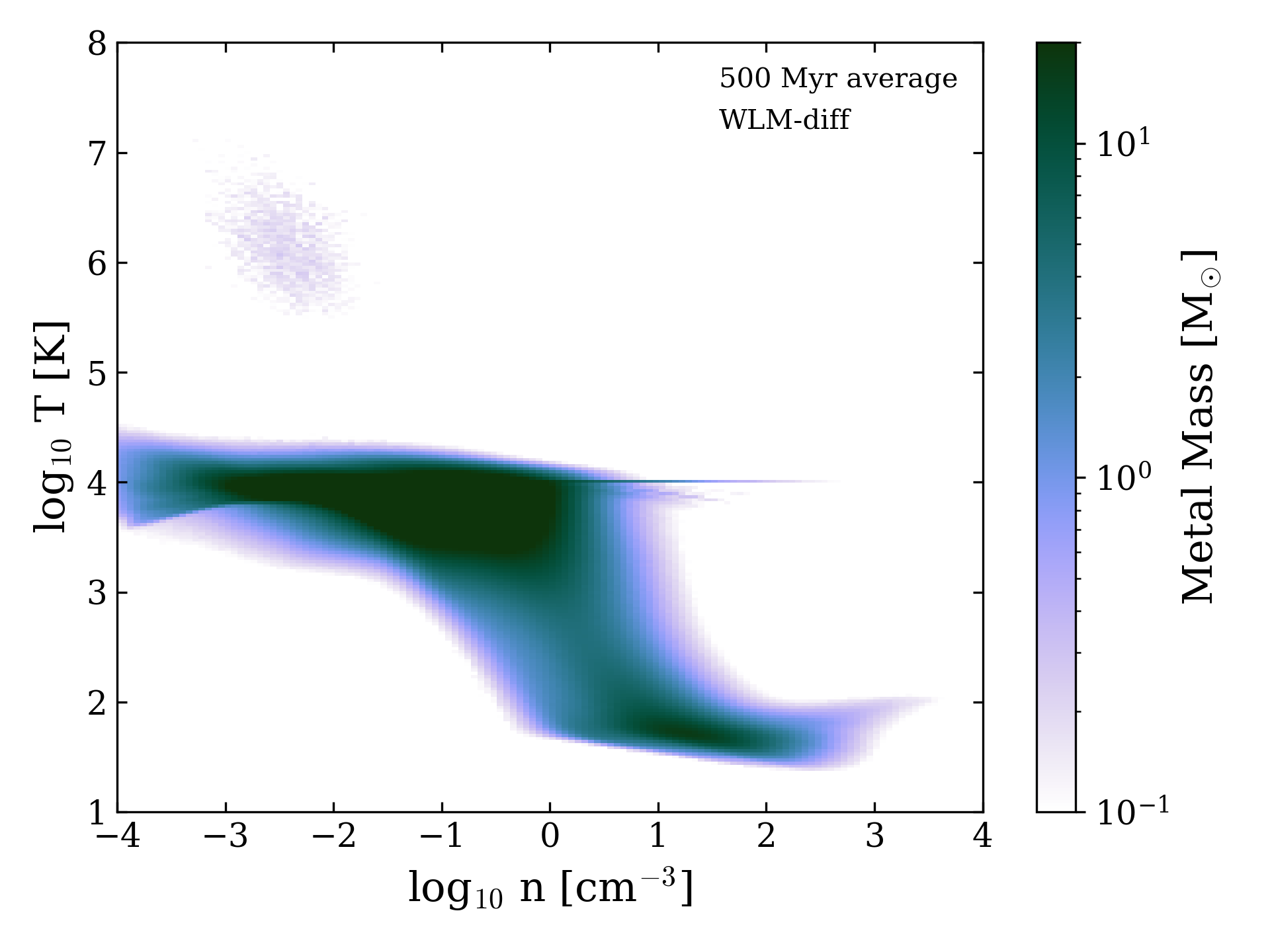}
    \caption{Density-temperature phase-space diagram, color coded by metal mass for the model WLM-nodiff (left) and the model WLM-diff (right). We note that in the run WLM-diff, we find that the apparent bifurcation in the metal mass is removed, in contrast with the run WLM-nodiff. Additionally, we find that there appears to be slightly more metal mass in the hot phase of the ISM, although the effect is minor (around 10 percent).}
    \label{fig:phase}
\end{figure*}

\begin{figure*}
    \centering
    \includegraphics[width=0.49\textwidth]{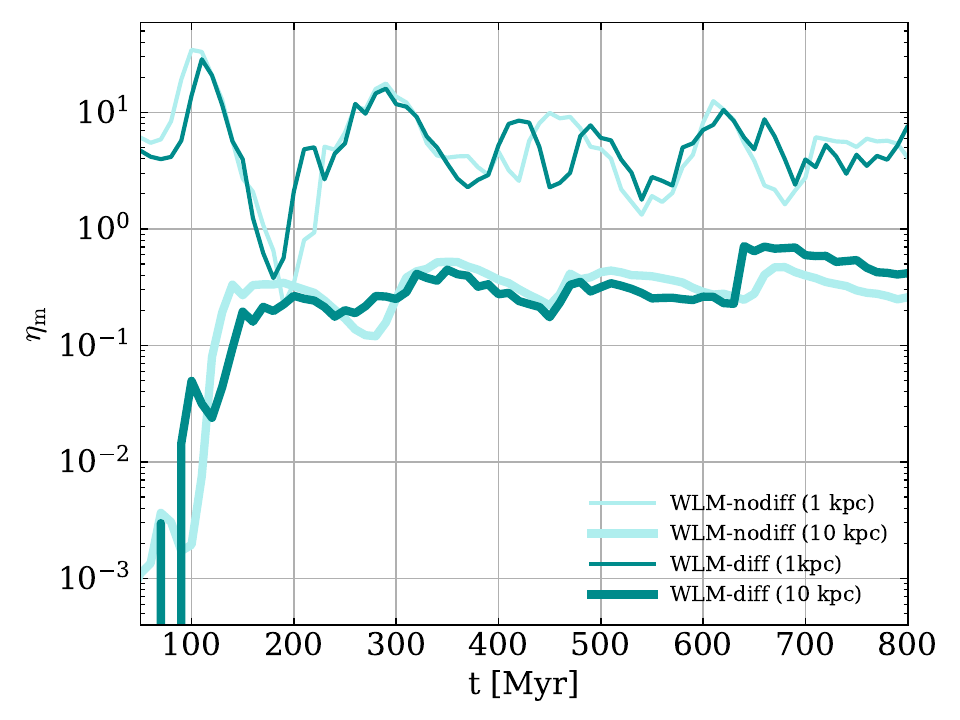}
    \includegraphics[width=0.49\textwidth]{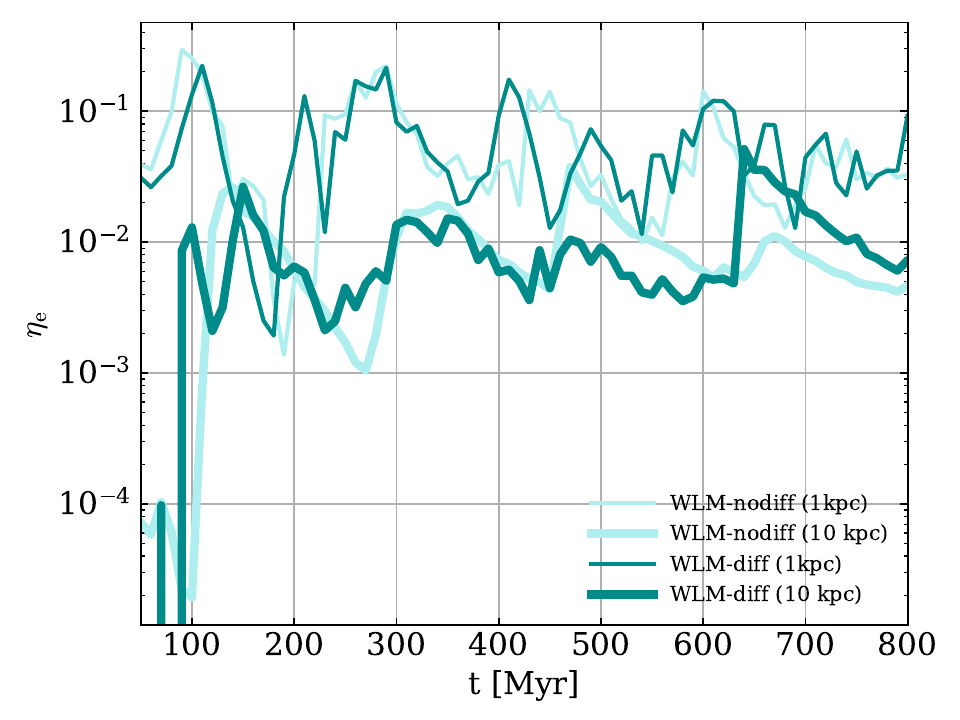}
    \caption{Time evolution of the mass loading factors (left) and energy loading factors (right) at heights of 1 kpc (thin solid lines) and 10 kpc (thick solid lines) for the models WLM-nodiff (light turquoise) and WLM-diff (dark turquoise). We find very little difference in the evolution of the global mass and energy loading factors at both low (1 kpc) and high (10 kpc) altitudes.}
    \label{fig:mass_energy_out}
\end{figure*}

\begin{figure*}
    \centering
    \includegraphics[width=0.49\textwidth]{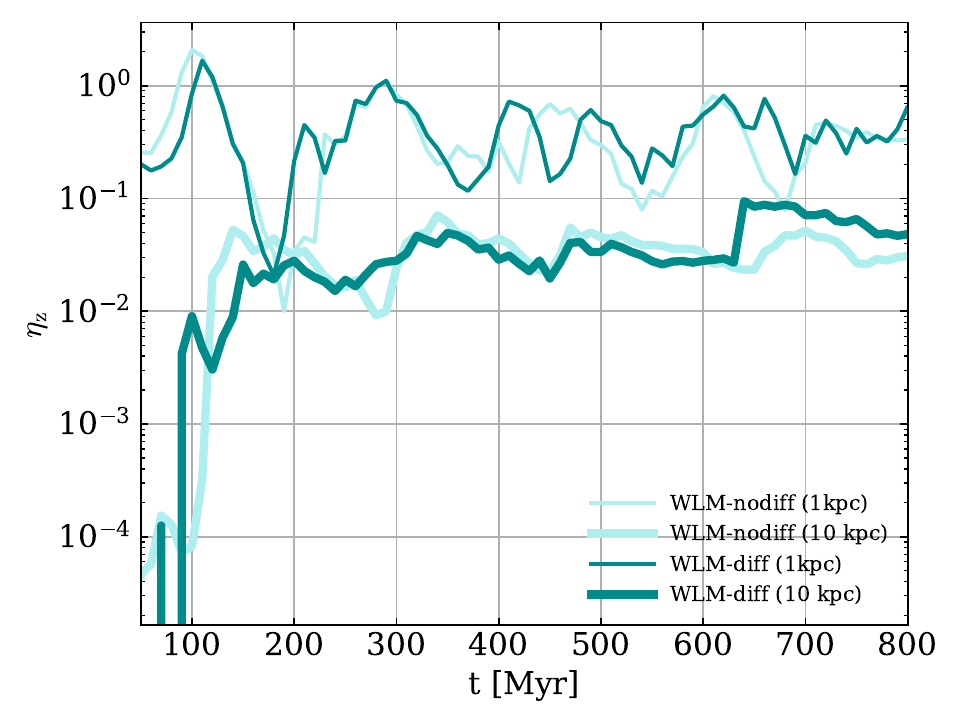}
    \includegraphics[width=0.49\textwidth]{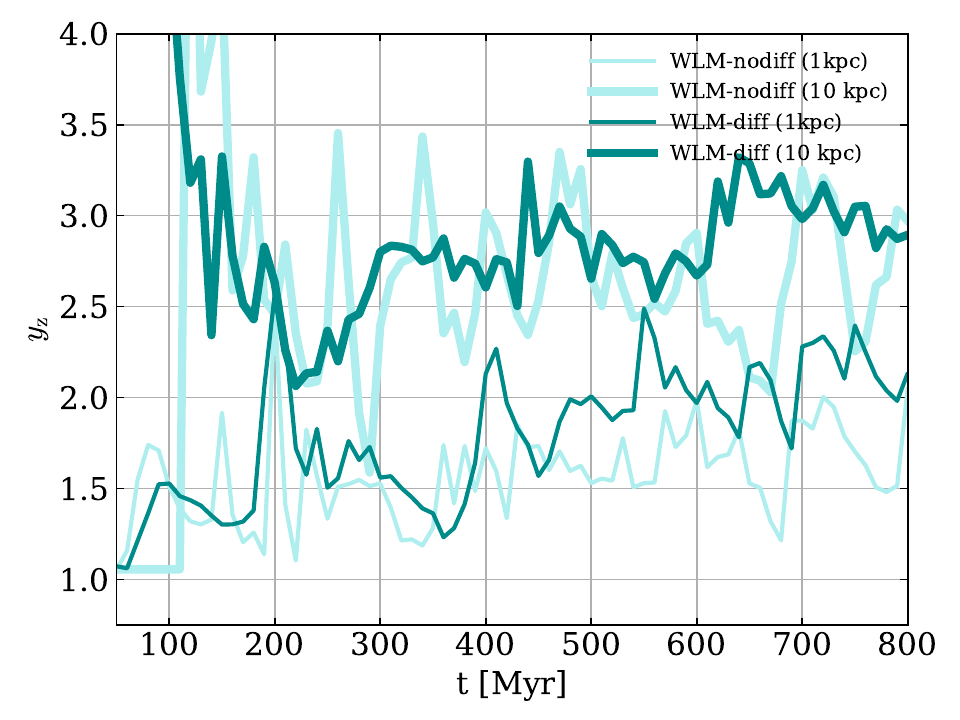}
    \caption{Time evolution of the metal loading factor (left) and the metal enrichment factor (right) at heights of 1 kpc (thin solid lines) and 10 kpc (thick solid lines) lines for the models WLM-nodiff (light turquoise) and WLM-diff (dark turquoise). The effect of the metal diffusion model on the metal loading factor is marginal. However, we see some slight increase in the metal enrichment factor in the WLM-diff run, with  a factor of up to 1.5 at low altitude (1 kpc) and up to 1.25 at high altitude (10 kpc). }
    \label{fig:metal_enrichement_out}
\end{figure*}

\begin{figure*}
    \centering
    \includegraphics[width=0.49\textwidth]{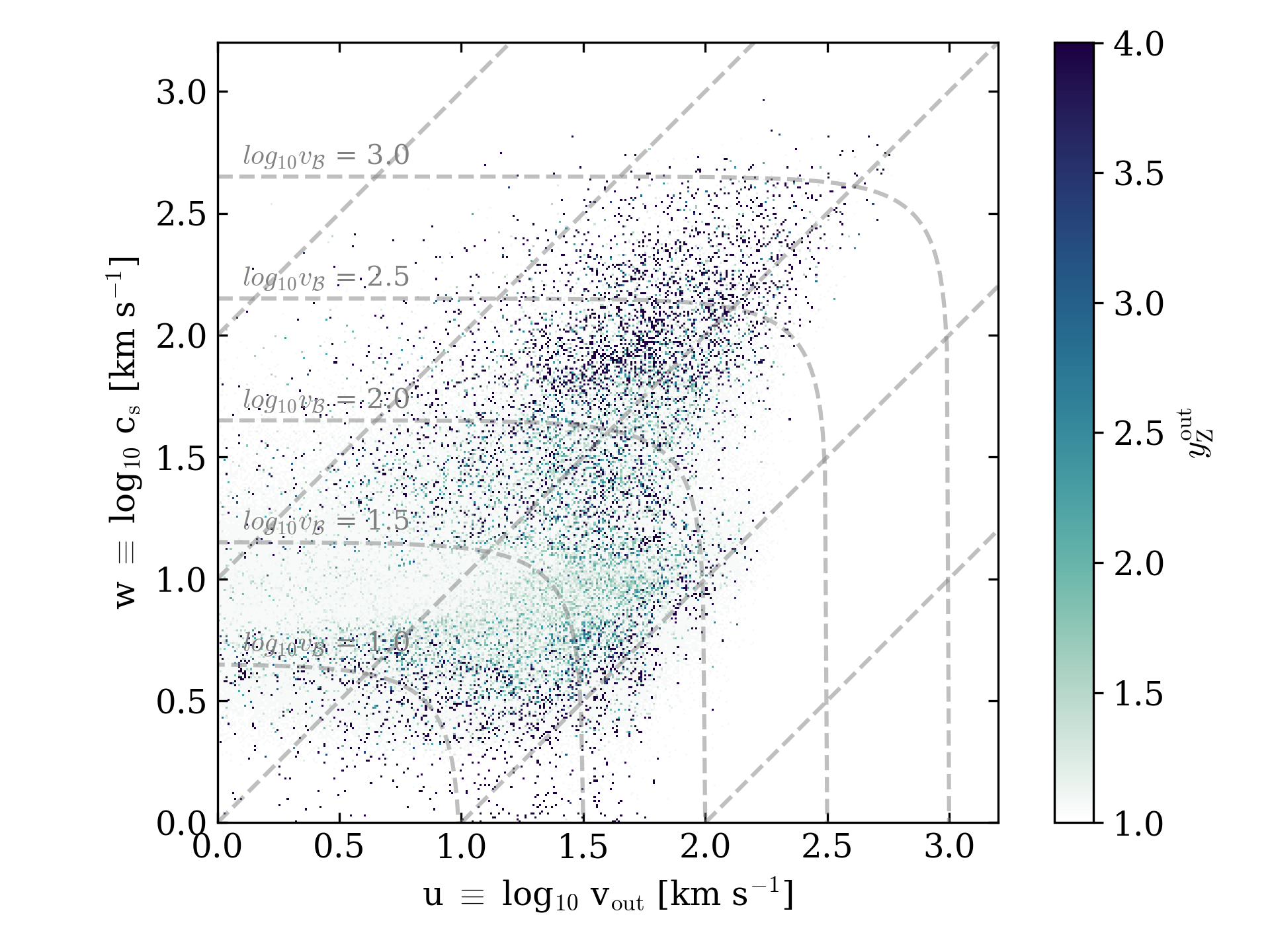}
    \includegraphics[width=0.49\textwidth]{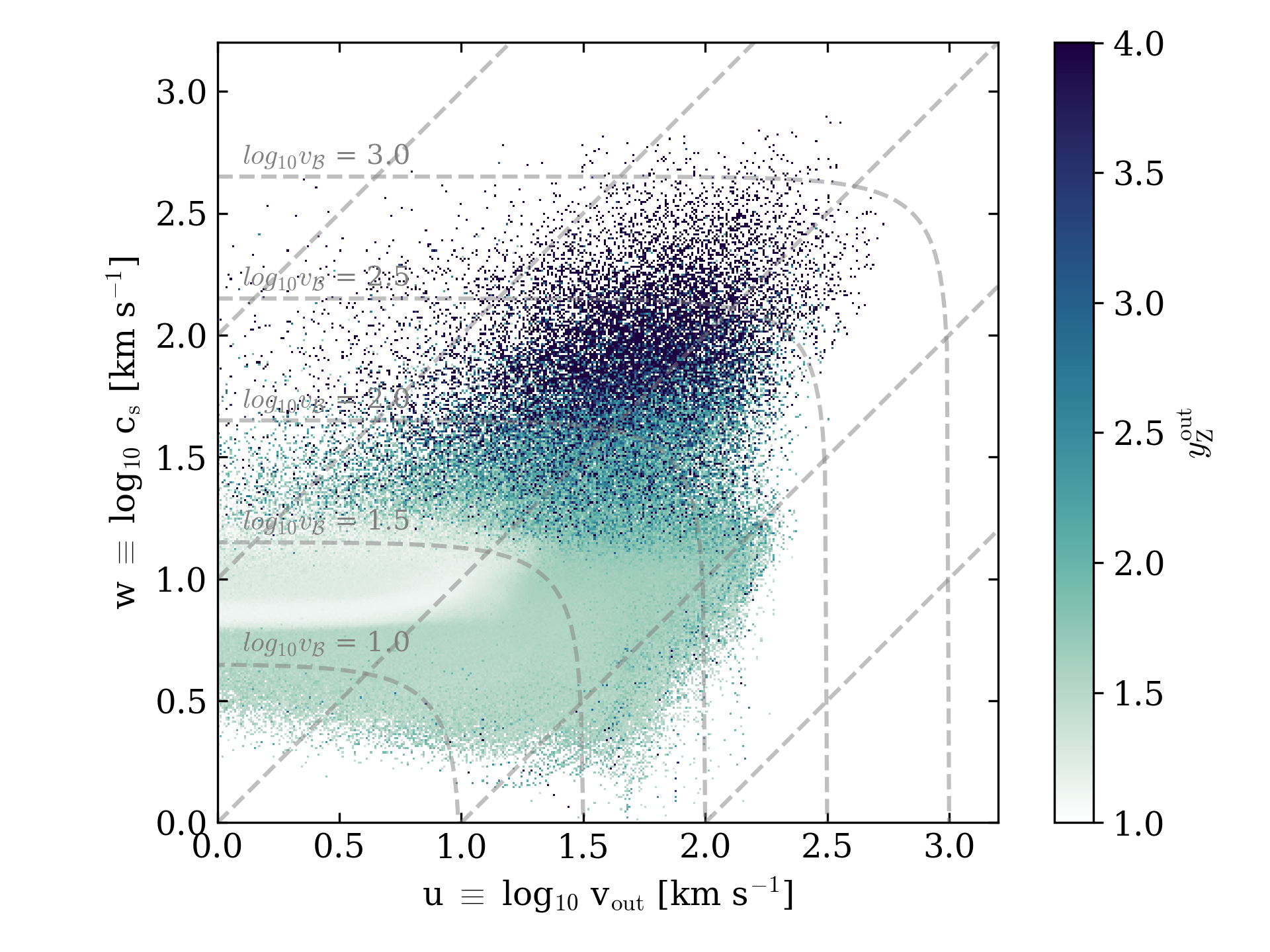}
    \includegraphics[width=0.49\textwidth]{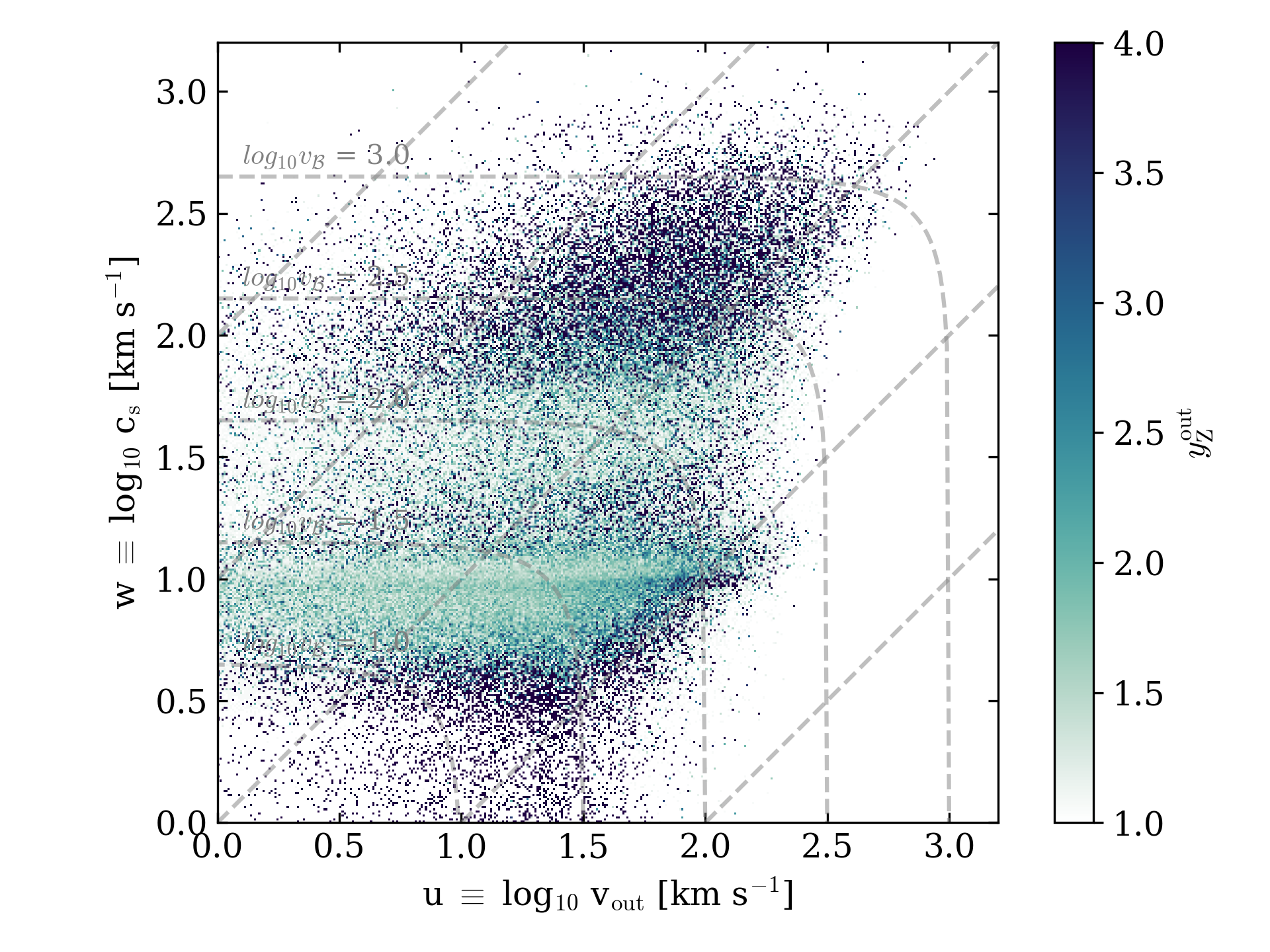}
    \includegraphics[width=0.49\textwidth]{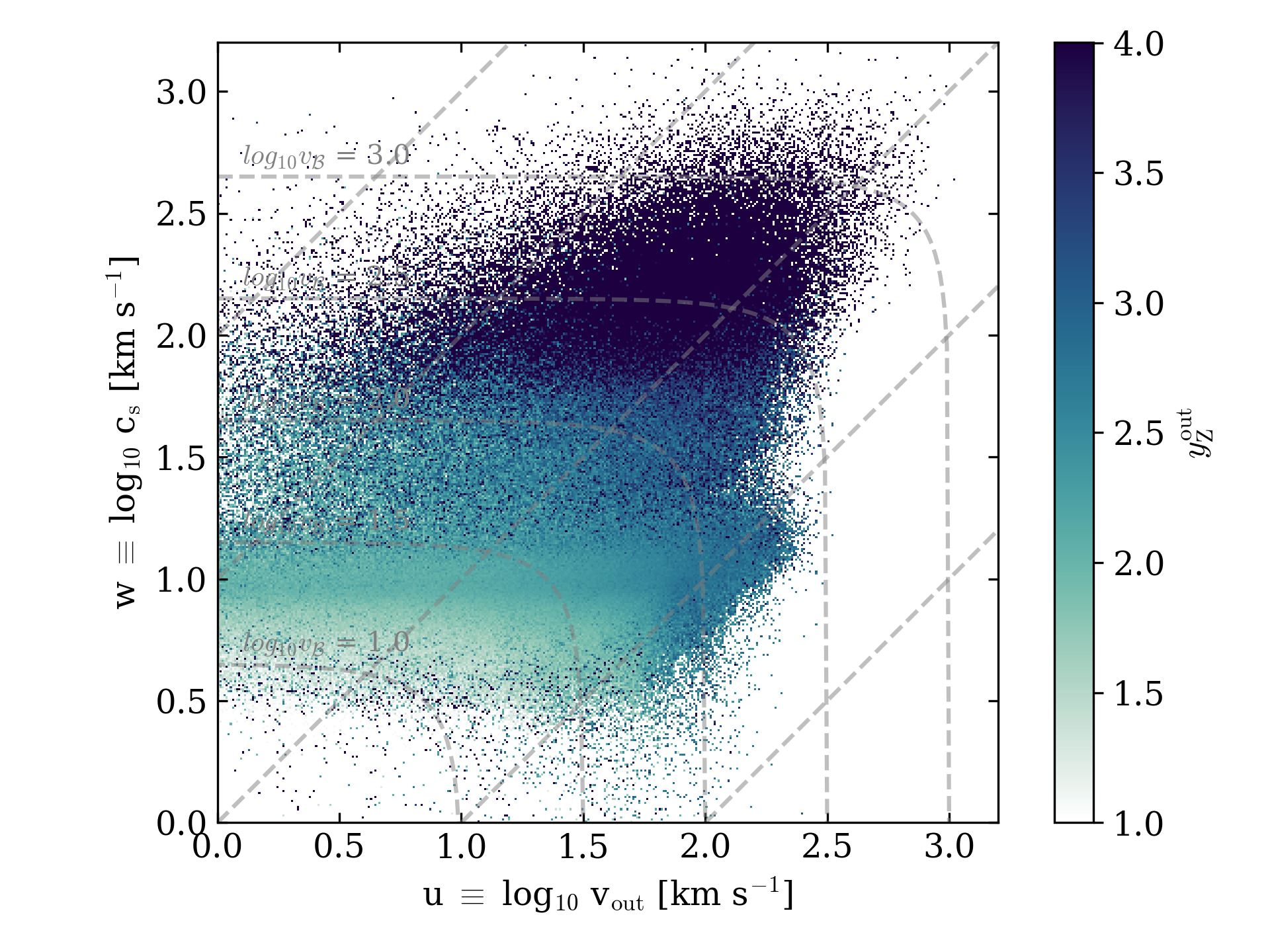}
    \caption{Joint 2D PDFs of outflow velocity and sound speed, color coded by the metal enrichment factor for the models WLM-nodiff (top left), WLM-diff (top right), LMC-nodiff (bottom left) and LMC-diff (bottom right). We find that the sound speed in the hot phase of the wind is higher by a factor of two when metal diffusion is enabled.}
    \label{fig:yout}
\end{figure*}

\begin{figure}
    \centering
    \includegraphics[width=0.49\textwidth]{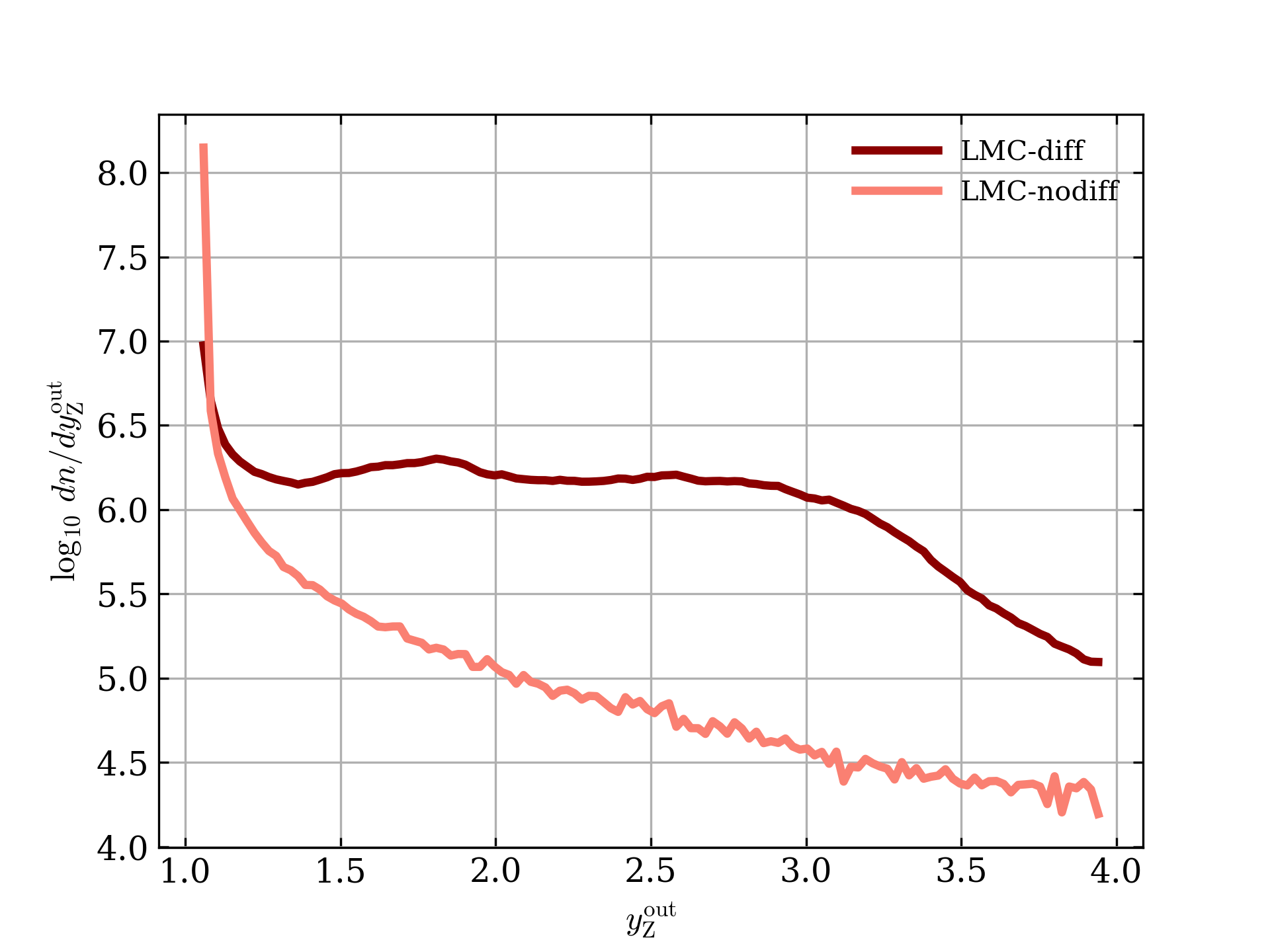}
    \caption{1D PDF of the distribution of the metal enrichment factor. }
    \label{fig:yout_1d}
\end{figure}

\begin{figure}
    \centering
    \includegraphics[width=0.49\textwidth]{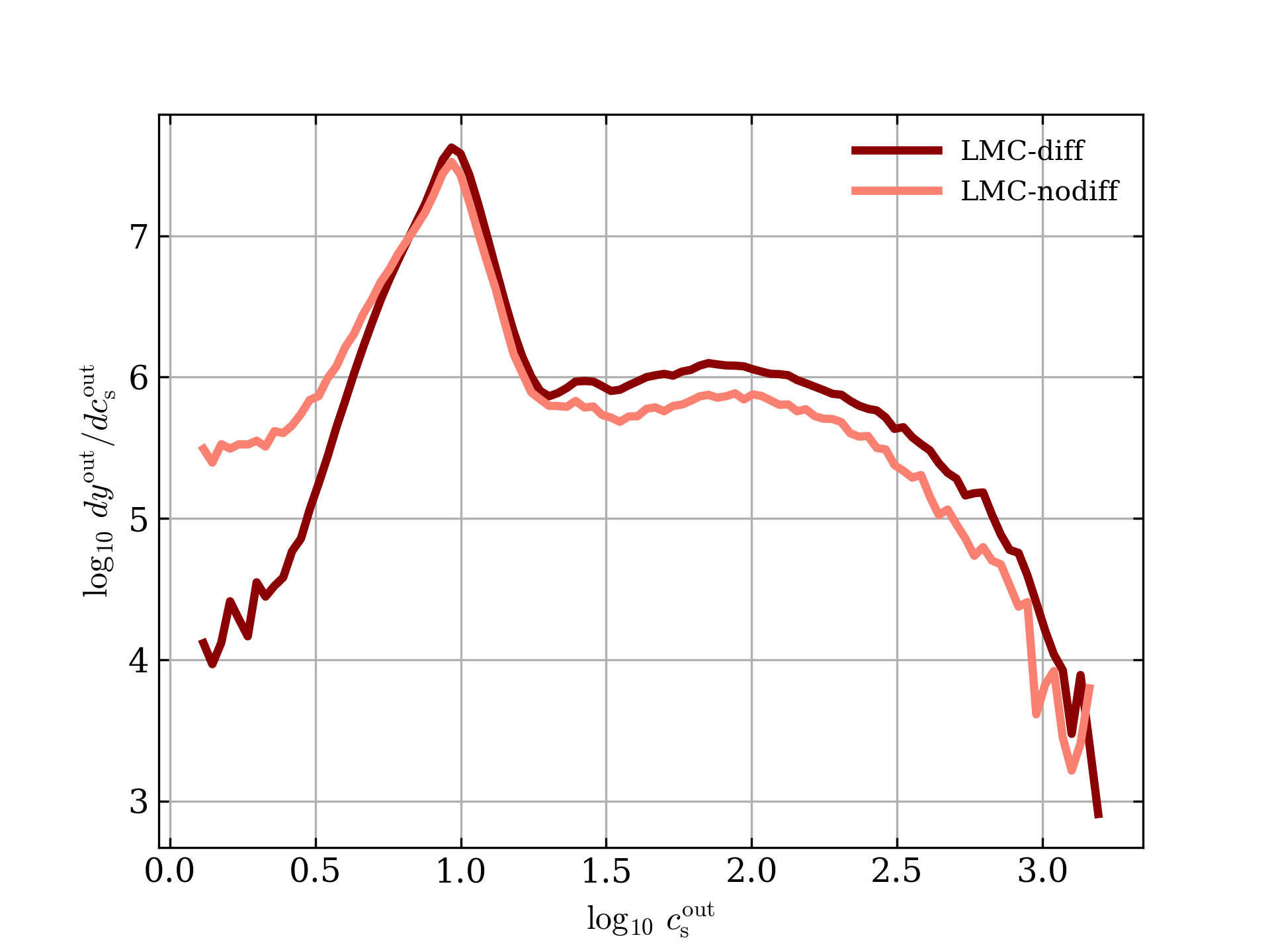}
    \includegraphics[width=0.49\textwidth]{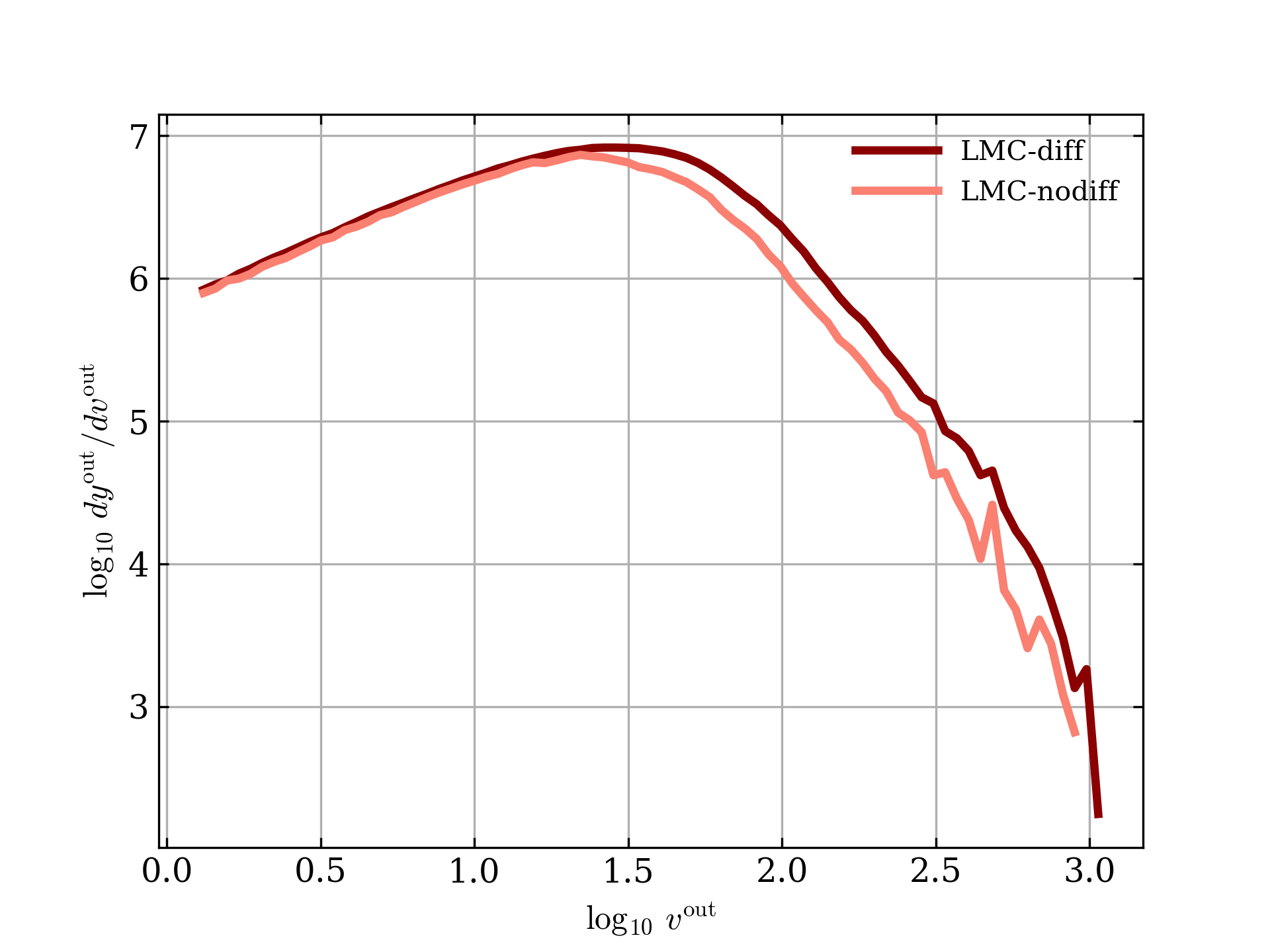}
    \includegraphics[width=0.49\textwidth]{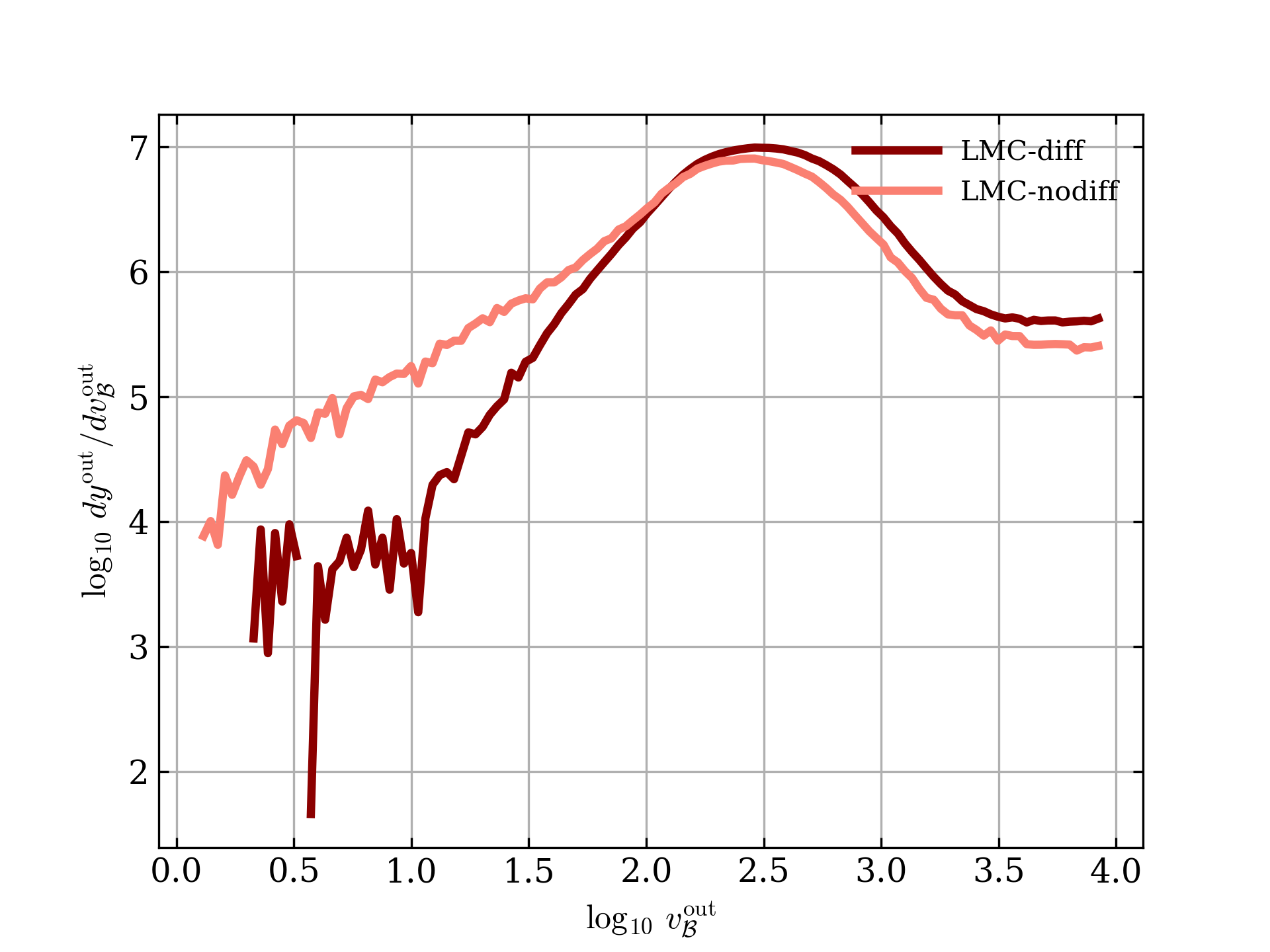}
    \caption{1D PDFs for the outflow velocity (top), the sound speed (center) and the Bernoulli-velocity (bottom) for the runs LMC-diff (dark red) and LMC-nodiff (salmon) weighted by y$^{\mathrm{out}}$.}
    \label{fig:1d_pdfs}
\end{figure}

\subsection{Definition of outflow loading factors}

We follow the definitions for outflow rate, mass loading, and enrichment factors of previous works -- specifically, the definitions presented in \citet{Hu2019}, \citet{Steinwandel2022} and \citet{SteinwandelLMC}. Generally, two important galactic wind quantities are mass-flux (m), and energy-flux (e), defined as:
\begin{align}
    {\bf \mathcal{F}}_\mathrm{m} = \rho \textbf{v},
\end{align}
\begin{align}
    {\bf \mathcal{F}}_\mathrm{e} = (\rho e_\mathrm{tot} + P)\textbf{v},
\end{align}
where $\rho$, $v$, and $P$ represent the fundamental fluid variables of density, velocity, and pressure, respectively, and $e_\mathrm{tot}$ is defined as the total energy per unit mass, $e_\mathrm{tot} = 0.5 \cdot \mathbf{v}^2 + u$, where $u$ is the internal energy per unit mass. In standard code units, $u$ takes the units km$^2$ s$^{-2}$. We note this to avoid confusion when converting these units to proper cgs units in the sections below. 

In this work, we measure the loading factors at a height of 1 kpc  above the disk in a slab 100 pc thick and at a height of 10 kpc  in a slab 1 kpc thick. 

We adopt the following definitions for outflow rates, with the total net flow rates defined as $\dot{M} = \dot{M}_\mathrm{out} - \dot{M}_\mathrm{in}$, $\dot{p} = \dot{p}_\mathrm{out} - \dot{p}_\mathrm{in}$ and $\dot{E} = \dot{E}_\mathrm{out} - \dot{E}_\mathrm{in}$. Outflow is defined by a positive value of the radial velocity: $\mathbf{v \cdot \hat{n}} = v_{r} > 0$ (this is actually v$_\mathrm{z}$)\footnote{We note that one could actually take proper radial velocity over the slab, which leads to a slightly higher value. However, that would be incorrect, as the radial velocity would have some angle compared to v$_\mathrm{z}$. One could correct that by offsetting the outflow rate computed over the radial velocity with the sine of the angle between the radial and the vertical velocity for a given particle. We tested all of these and note that they differ by only a small percentage. The reason for this is that since rotational velocity is low in dwarfs, the vertical velocity will always dominate the radial velocity contribution at a given height. We expect this to change for larger galaxies such as the Milky Way that are more strongly rotationally supported.}. Hence, we can write down the discrete outflow rates for mass and energy:
\begin{align}
    \dot{M}_\mathrm{out} = \sum_{i, v_{i,r} > 0} \frac{m_{i}v_{i,r}}{dr},
\end{align}
\begin{align}
    \dot{E}_\mathrm{out} = \sum_{i, v_{i,r} > 0} \frac{m_{i} [v_{i}^2 + \gamma u_{i}]v_{i,r}}{dr},
\end{align}
where $m_i$ is the mass and $v_i$ is the velocity of the gas elements in the slab, and $dr$ is the thickness of the slab. We use $P = \rho u (\gamma -1)$ as an equation of state. We compute metal outflow rates using:
\begin{align}
    \dot{M}_\mathrm{out, Z} = \sum_{i, v_{i,r} > 0} \frac{Z_{i} m_{i}v_{i,r}}{dr},
\end{align}
where $Z_{i}$ is the metallicity of particle $i$.
In this paper, we investigate loading factors that are obtained by normalizing to a set of reference quantities for mass ($\eta_\mathrm{m}$), metals ($\eta_\mathrm{Z}$), and energy ($\eta_\mathrm{e}$):
\begin{enumerate}
    \item outflow mass loading factor: $\eta_\mathrm{m}^\mathrm{out} = \dot{M}_\mathrm{out} / \overline{\mathrm{SFR}}$,
    \item energy outflow loading factor: $\eta_\mathrm{e}^\mathrm{out} = \dot{E}_\mathrm{out}/(E_\mathrm{SN} R_\mathrm{SN})$,
    \item outflow metal loading factor: $\eta_\mathrm{Z}^\mathrm{out} = \dot{M}_\mathrm{Z, out} / (m_\mathrm{Z} R_\mathrm{SN})$,
\end{enumerate}
where we adopt $E_\mathrm{SN} = 10^{51}$ erg, $p_{\rm SN} = 3 \cdot 10^{4}$ M$_{\odot}$ km s$^{-1}$, supernova rate $R_{\rm SN} = \overline{\mathrm{SFR}}/(100 \; {\rm M}_{\odot})$ and (IMF-averaged) metal mass  $m_\mathrm{Z} = 2.5 \; {\rm M}_{\odot}$, which we take for best comparison with previous dwarf galaxy studies from \citet[][]{Hu2019}.
We compute the mean star formation rate $\overline{\mathrm{SFR}}$ over a 500-Myr time span for WLM-diff and WLM-nodiff that ranges from 300 Myr to 800 Myr. For LMC-diff and LMC-nodiff, we average from 200 to 500 Myrs (a time span of 200 Myrs). We follow this procedure for LMC because the total runtime of the original run LMC-nodiff, as published in \citet{SteinwandelLMC}, is only 500 Myrs and the new run (in this work) labeled  LMC-diff has been restarted from a snapshot of LMC-nodiff at time $t=400$ Myr. We note that the mean star formation rate in those cases is computed by taking the mean of the star formation rate over 200 snapshots for LMC-diff and LMC-nodiff and 500 snapshots for WLM-diff and WLM-nodiff. The star formation rate for each snapshot is computed by summing the masses of the young stellar population with stellar ages younger than 40 Myr.

Additionally, we define metal enrichment factors that quantify the degree to which metals are over- or under-abundant in galaxy outflows relative to the ISM. This can be achieved by normalizing the metal outflow rate to the mass outflow rate, weighted by the background metallicity of the galactic ISM, $Z_\mathrm{gal} = 0.1 \; Z_{\odot}$: 
\begin{enumerate}
    \item outflow enrichment factor: $y_\mathrm{Z}^\mathrm{out} = \dot{M}_\mathrm{Z, out}/ (\dot{M}_\mathrm{out} Z_\mathrm{gal})$, 
    \item inflow enrichment factor: $y_\mathrm{Z}^\mathrm{in} = \dot{M}_\mathrm{Z, in}/ (\dot{M}_\mathrm{in} Z_\mathrm{gal})$.
\end{enumerate}

\subsection{Diffusion vs. turbulent mixing}

In Fig.~\ref{fig:split_diff}, we show the mass-weighted joint 2D PDFs of the diffusion time vs. the turbulent mixing timescale of the system (mixing by advection). The diffusion timescale is computed via: 
\begin{align}
    t_\mathrm{diff} = \frac{h^2}{D},
\end{align}
where $h$ is the radius of compact support of our cubic spline kernel and $D$ is the diffusion coefficient computed as in eq.~\ref{eq:diff}. The mixing timescale by advection is computed via: 
\begin{align}
    t_\mathrm{mix} = \frac{h}{\sigma},
\end{align}
where $\sigma$ is the kernel-averaged velocity dispersion. We show this distribution for the full phase space (top left); the cold neutral medium (CNM, top right), which we define with a simple temperature cut of gas below 300 K; the warm neutral medium (WNM, bottom left), with gas in the temperature range between 300 and 8000 K; and the warm ionized medium (WIM, bottom right), with gas between 8000 and 500,000 K. We generally find that most of the gas mass is dominated by our diffusion model, apart from some gas in the diffusive hot phase, where it appears that turbulent mixing is more efficient than diffusion.

\subsection{Star formation rate and ISM structure}

In Fig.~\ref{fig:sfr}, we show the star formation rate of all of our simulations: WLM-nodiff (light turquoise), WLM-diff (dark turquoise), LMC-nodiff (salmon) and LMC-diff (dark red). We note that due to the computational expense of the models LMC-nodiff and LMC-diff, we decided not to carry out the simulation LMC-diff from scratch but rather to restart it from a snapshot of the model LMC-nodiff at 400 Myrs of evolution. This is roughly one wind driving cycle for this model and we do believe that this is sufficient for the time-averaged wind quantities discussed below. However, for the time evolution of the loading factors, 100 Myr of evolution has limited predictive power. Thus we probe the long-term evolution of the impact of metal diffusion on star formation history with the WLM-nodiff and WLM-diff runs. Generally, we find a weak effect of metal diffusion on star formation history. However, we do note that the time-averaged star formation rate between 150 Myrs and 800 Myrs for the model WLM-diff is about a factor of 1.3 higher than the star formation rate for the model WLM-nodiff. This is consistent with the restart of the model LMC-diff, which also shows a higher star formation rate (by a factor of about 1.5) than the model LMC-diff (see the summary in Table~\ref{tab:loadings}). We do note that in LMC-diff and LMC-nodiff, this might be influenced by restarting from a snapshot where positions and velocities have been stored in single precision. Regardless, the  higher average star formation rate in WLM-diff compared to WLM-nodiff is a fact. 

In Fig.~\ref{fig:phase}, we show the time averaged phase-space diagrams over 500 Myrs of density and temperature weighted by metal mass for the model  WLM-nodiff (left) and the model WLM-diff (right). The impact of the metal diffusion model on these quantities is weak, but there are some notable differences we would like to highlight. First, the hot phase of the ISM is only weakly impacted by the presence of metal diffusion, although it appears that there is slightly more metal mass in the hot ISM phase with metal diffusion.  Additionally, with metal diffusion included, more metal mass appears in the cold dense part of the ISM and the phase-space diagram is smoother. Specifically, the feature that appears between 10 and 100 K and 0.1 and 1 cm$^{-3}$ in the  left panel of Fig.~\ref{fig:phase} (where a lot of metal mass exists in the run WLM-nodiff) is absent from the run WLM-diff. This gas in the nodiff run originates from the fountain of previous outflow events with very high metallicity: it has relatively low outflow velocity, is quickly reaccreted and cools very rapidly out of the hot ISM phase. These events have much less impact in the presence of metal diffusion in the run WLM-diff, and remove much of this relatively diffuse, metal-rich cold gas.

\subsection{Time evolution of loading factors}

In Fig.~\ref{fig:mass_energy_out}, we show the time evolution of the mass loading factors (left) and energy loading factors (right) at heights of 1 kpc (thin solid lines) and 10 kpc (thick solid lines) for the models WLM-nodiff (light turquoise) and WLM-diff (dark turquoise). At the lower altitude, we find a mass loading factor that is consistently between 1 and 10, which is comparable to the findings of previous studies of similar models at that height \citep[][]{Hu2016, Hu2017, Hu2019, Hu2023, Gutcke2021, Smith2021, Steinwandel2022, SteinwandelLMC, Hislop2022, Lahen2023}. However, we do note that even at the higher altitude of 10 kpc, we still find a rather strongly mass-loaded wind with a mass loading factor between 0.1 and 1. This is in line with the results of \citet{Hu2019},  who used the pressure-energy implementation of SPH in a previous version of our code \textsc{gadget} and whose  values at the same height  are  larger than ours by  a factor of about 2-3. The discrepancy might be explained by the fact that while we measure in plane-parallel slabs at that height, \citet{Hu2019} measures in a spherical shell at the same height. This might explain the difference of a factor of two. We do note that our values are significantly larger than those reported by \citet{Smith2021} for a similar model with a similar physics implementation in the code \textsc{arepo}. We will discuss possible reasons for this in  greater detail in Section \ref{sec:discussion}. For the energy loading, we find values of 0.1 at height 1 kpc and around 0.01 at height 10 kpc. These are, again, quite comparable to the values reported by \citet{Hu2019} but much higher than those reported by \citet{Smith2021} for similar systems. The  metal diffusion does not significantly affect the energy loadings. 

In Fig.~\ref{fig:metal_enrichement_out}, we show the time evolution of the metal loading factor (left) and the metal enrichment factor (right) at heights of 1 kpc (thin solid lines) and 10 kpc (thick solid lines) for the models WLM-nodiff (light turquoise) and WLM-diff (dark turquoise). We find only a little difference (below the 10 percent level) in the modeling with and without metal diffusion. This is only slightly higher than the difference between runs, which is around 5 percent. The metal loading at a height of 1 kpc is between 0.1 and 1. At a height of 10 kpc, we find lower values for the metal loading, between 0.01 and 0.1. Again, these values are quite comparable to those in \citet{Hu2019}. This comparison is particularly important for us since those simulations employ a very similar physics model along with metal diffusion. At the lower altitude of 1 kpc, we find that the metal enrichment factor is between 1.25 and 1.5 for the model WLM-nodiff and between 1.5 and 2.0 for the model WLM-diff. Thus we do find that there is some minor impact on the metal enrichment at low altitude. However, since the star formation rate in WLM-diff is higher, on average, by a factor of 1.3, it is possible that this difference originates from the slightly higher average star formation rate that is used to obtain these loading factors. Nevertheless, the higher star formation rate in the model WLM-diff  likely stems from more efficient cooling, due to more efficient metal mixing in the ISM, in the phase prior to the launching of galactic winds by superbubbles. The metal enrichment factor is larger, between 2 and 3, at the higher height of 10 kpc. While the numerical values between the two models are quite comparable, the metal enrichment history of the model WLM-nodiff appears more bursty. That might indicate that cooling losses due to metal line cooling in the hot wind are more dominant when there is no metal diffusion model implemented.  

Additionally, we show the 1D-PDFs for dn/dy$^{\mathrm{out}}$ for the runs LMC-diff (darkred) and LMC-nodiff (salmon) in Fig.~\ref{fig:yout_1d}. We find that the strong gradient in the metal enrichment factor that is present in the run LMC-nodiff is removed in the run LMC-diff. This can be understood as the metal diffusion model driving the system to an equilibrium state in which the metals are more uniformly distributed within the outflow. 

In Fig.~\ref{fig:1d_pdfs} we show the 1D PDfs for the outflow velocity, the sound speed, and the Bernoulli-velocity which is defined via:

\begin{align}
    v_\mathcal{B} = \left(v_\mathrm{out}^2 + \frac{2 \gamma}{\gamma -1} c_\mathrm{s}^2\right),
\end{align}

\subsection{Phase structure of the outflows}

In Fig.~\ref{fig:yout}, we show the joint 2D PDFs of outflow velocity and sound speed, color coded by the metal enrichment factor for the models WLM-nodiff (top left), WLM-diff (top right), LMC-nodiff (bottom left) and LMC-diff (bottom right), measured at 1 kpc height. These are time averaged, in the case of the LMC models over 100 snapshots between 400 and 500 Myr of evolution, in the case of the WLM models over 500 Myrs between 300 and 800 Myrs of evolution. The metal enrichment factor is the quantity that is most dramatically affected by the presence of a metal diffusion model; this is the key finding of this study. While the mass, metal and energy budgets of the outflow are only weakly affected by a metal diffusion model, the metal enrichment factor is different. 

We find that, in the absence of a metal mixing model for the WLM-nodiff and LMC-nodiff runs, there is a complete phase-space separation between the metal-enriched hot outflow and the metal-under-enriched warm fountain. And so we find that without metal diffusion,  most of the outflow is composed of ambient gas, with only a few cells of very high metallicity (up to $20$ Z$_{\odot}$) in the hot outflow. We point out that the gap in the distribution at a sound speed of around 40 km s$^{-1}$ corresponds to a temperature of $\sim$$10^5$ K. This gap originates due to the high metallicity gas being able to cool rapidly as this is where the cooling curve peaks and the hot gas can cool more efficiently down to the low temperature regime.  However, they do not mix effectively with the warm outflow and instead cool to very low temperatures. This feature disappears in the runs with metal diffusion (WLM-diff and LMC-diff), where it becomes clear that the highly metal-enriched part is quickly mixing with the not enriched ambient gas at a height of around 1 kpc  above the midplane, leading to a gradient of the metal enriched factor that is consistent with previous simulations where metal mixing is included in Eulerian codes, due to the intrinsic numerical diffusion of the Riemann solver \citep[e.g.,][]{Kim2020}.

\section{Discussion}
\label{sec:discussion}

\begin{table*}
    \caption{Summary of the mean loading factors once the systems have reached quasi-self-regulating states at heights of 1 kpc and 10 kpc above the midplane. These are averaged between 300 Myr and 800 Myr for WLM-diff and WLM-nodiff and 300 to 500 Myrs for LMC-diff and LMC-nodiff. The mean star formation rate is computed over the same time period for these models.} 
    \label{tab:loadings}
    \begin{tabular}{cccccccccccc}
      \hline
	  Name & SFR [M$_{\odot}$ yr$^{-1}$] & $\eta_\mathrm{M}^\mathrm{out, 1}$ & $\eta_\mathrm{Z}^\mathrm{out, 1}$ & $\eta_\mathrm{E}^\mathrm{out, 1}$ & $y_\mathrm{Z}^\mathrm{out, 1}$ & $\eta_\mathrm{M}^\mathrm{out, 10}$ & $\eta_\mathrm{Z}^\mathrm{out, 10}$ & $\eta_\mathrm{E}^\mathrm{out, 10}$ & $y_\mathrm{Z}^\mathrm{out, 10}$\\
	  \hline
      WLM-nodiff & 0.0016 & 6.48 & 0.27 & 0.059 & 0.0084 & 0.40 & 0.03 & 1.55 & 2.55\\
      WLM-diff & 0.0021 & 5.91 & 0.29 & 0.058 & 0.00089 & 0.41 & 0.033 & 1.77 & 2.54\\
      LMC-nodiff & 0.049 & 2.2 & 0.5 & 0.1 & 2 & 0.22 & 0.05 & 0.02 & 3.5\\
      LMC-diff & 0.075 & 2.7 & 0.6 & 0.1 & 2.1 & 0.27 &  0.06 & 0.025 & 3.7\\
      \hline
      \end{tabular}
\end{table*}

\subsection{Comparison with previous work}

The most relevant studies to this work are \citet{Hu2019}, \citet{Smith2021} and \citet{Gutcke2021}, as they employ similar models for the treatment of the interstellar medium (ISM) in WLM-type dwarf galaxies. 
\citet{Hu2019} carried out an inverse resolution study of a WLM-type dwarf with a star formation rate of around $10^{-4}$ M$_{\odot}$ yr$^{-1}$, with four simulations at 1 M$_{\odot}$, $5\times 10^{-4}$ M$_{\odot}$, $25 \times 10^{-4}$ M$_{\odot}$ and $125 \times 10^{-4}$ M$_{\odot}$ in terms of mass resolution, to study the convergence of radial profiles of loading factors for mass, energy, momentum and metals. Our results for WLM-nodiff and WLM-diff are in good agreement with these results of \citet{Hu2019}. However, we find a mass loading at a height of 1 kpc that is a factor of 3-4 lower than in their simulations. The reason for this is that the star formation rate in the simulations of \citet{Hu2019} is  smaller than in our system by a factor of around 10. Hence, their mass loading at a height of 1 kpc is about 30. This is consistent with the simulations of \citet{Steinwandel2022} who find similar values for a lower-surface-density version of WLM that has been adopted in previous simulations. At the higher altitude of 10 kpc, our loading factors are consequently  again lower than those reported by \citet{Hu2019}. Despite the difference in star formation rate, the values for energy loading at 1 kpc and at 10 kpc are very comparable to the results of \citet{Hu2019}, where we find values of around 0.1 at 1 kpc and around 0.03 at 10 kpc.  The metal loading is similar as well, where \citet{Hu2019} reports values of 0.2 and we find values between 0.05 (10 kpc) and 0.5 (1 kpc). These values are not sensitive to the use of a sub-grid metal mixing model. 

\citet{Smith2021} carried out a large suite of over 30 simulations of a WLM-like system with the moving mesh code \textsc{arepo}, very similar to our WLM-nodiff and WLM-diff runs. While they find relatively similar values for the mass and energy loading at 1 kpc, they find almost no mass- or energy-loaded outflow at a height of 10 kpc in their most comparable run with SNe, photoelectric heating and photoionizing radiation. Only in their SN-only run do they find comparable values to ours for the mass and energy loading at 10 kpc, with values ranging from 0.1 to 10 for the mass loading and from 0.01 to 1 for the energy loading. It is quite interesting that in their simulations, the outflow appears to be completely quenched with SNe, photoelectric heating and photoionizing radiation. The origin of this is not quite clear. There are a few differences between their simulations and ours. First, their simulations are of slightly lower resolution (20 M$_{\odot}$) when compared to ours at face value (4 M$_{\odot}$), and \citet{Smith2021} adopted \textsc{grackle} for the cooling while we follow the non-equilibrium networks of \citet{Nelson1997}, \citet{Glover2007a}, \citet{Glover2007b} and \citet{Clark2012}. Thus, it might a combination of these factors that causes the difference in outflow at 10 kpc. They did not look into the metal loading in their simulations, so this comparison cannot be made. 

\citet{Gutcke2021} carried out a set of three simulations of the low-surface-density WLM version that has been studied in greater detail in \citet{Hu2017}, \citet{Hu2019} and \citet{Steinwandel2022} with the moving mesh code \textsc{arepo}. This system has a star formation rate that is lower by roughly an order of magnitude than the star formation rate in our system. However, one of their model variations includes variable SN energies following the yield of \citet{Sukhbold2016}, which appears to have a similar star formation rate to our dwarf. They measure their loading factors at 2 kpc and find values for their mass loading between 1 and 10 and for their metal loading between 0.1 and 1. These values are not too dissimilar from what we report at 1 kpc height. The choice to measure at 2 kpc is interesting because our simulations and the simulations of \citet{Hu2019} indicate that this is the height of ``hot wind formation,'' marking the point where the hot wind starts to dominate the energetics of the wind as the warm fountain flow drops out and is recreated. If this is similar for the simulations of \citet{Gutcke2021}, then the values at 2 kpc that they measure, specifically for mass and metal loading, should not differ too much from the values that we measure at 1 kpc. They do not report their loading factors at higher altitude, so no meaningful comparison to our values at 10 kpc can be made. It is interesting to note that they find a rather low energy loading, between $10^{-4}$ and $10^{-3}$, even at 2 kpc. This is lower than our energy loading of around 0.01-0.05 measured at the height of 10 kpc above the midplane.

It is worthwhile to note that both of these sets of simulations with \textsc{arepo} intrinsically adopt metal diffusion via the mass-flux from cell to cell.

\citet{Fielding2017} ran a similar global setup for dwarf galaxies in a leaner setup without self-gravity and cooling; they report values for the mass and energy loading of around 1 and 0.01, respectively. These are measured roughly around the launching radius and are smaller than the values we find at 1 kpc, which should be a good proxy for the values at the launching radius. No metal loading has been reported for these simulations.

\citet{Li2017} ran ``Tallbox'' simulations with the grid code \textsc{enzo}. For mass, metal and energy loading, they report the numbers 6, 0.7 and 0.2, respectively. These are in very good agreement with the values that we find at a height of 1 kpc. As these are Eulerian simulations, they do adopt mass transfer between cells and metals are spread by numerical diffusion between neighboring cells.

One explicit metal diffusion model that is relevant to compare to our simulations is presented in \citet{Shin2021}, who simulated a Milky Way-like isolated galaxy with the ICs from the \textit{Agora} project \citep{Kim2014_agora, Kim2016_agora} with the codes Gadget-3 \citep[e.g.,][]{springel05} and Gizmo \citep[e.g.,][]{Hopkins2015}. They essentially adopted the same metal diffusion model as the one that we implemented in our version of Gadget-3, and overall, their conclusions are similar to ours.  In particular, they also report a minor effect of metal diffusion on galaxy evolution. However, the overall trends appear to be the same, since \citet{Shin2021} report a higher star formation rate with metal diffusion. The trend seems to be weaker for us, but the time averaging in our simulations shows that the mean star formation rate in the ``diff'' runs is higher by a factor of 1.3 to 1.5 than in the ``no-diff'' runs. However, \citet{Shin2021} perform only a limited investigation of galactic outflows and a one-to-one comparison to their work is tricky. Nevertheless, they do report slightly higher average outflow velocities, which we can infer as well from our phase-diagrams in Fig.~\ref{fig:yout}. However, we attribute this  to the  higher overall level of star formation activity (especially in the run LMC-diff), which might  in turn be related to the more efficient cooling due to the metal mixing in the dense ISM.

\subsection{The need for metal diffusion models in Lagrangian codes}

The question remains: ``Do we need a metal diffusion model in simulations of galaxy evolution?'' The answer to that question is ``It depends.'' Considering Fig.~\ref{fig:sfr} and the values for the mean star formation rate in the four runs that we have (see Table \ref{tab:loadings}), we do find that the star formation rate is about a factor of 1.5 larger when metal diffusion is included. That trend is modest but measurable across run-to-run variations of our WLM-like systems. Hence, this effect is supported by the evidence we present. This goes hand in hand with the minor impacts on the distribution of the metal mass that we show in Fig.~\ref{fig:phase}. Again, these effects are modest, but there is no reason why we should prefer the ``nodiff'' runs over the ``diff'' runs in this respect. The overall effect on mass and energy loading factors, as reported in our Fig.~\ref{fig:mass_energy_out}, Fig.~\ref{fig:mass_out} and Fig.~\ref{fig:energy_out}, seems to be very minor. This statement does not apply to the height that we measure at (1 or 10 kpc). There are two separate effects related to the metal enrichment in the right-hand panel of Fig.~\ref{fig:metal_enrichement_out}. We find  that at later times, the metal enrichment in the run WLM-diff at 1 kpc is increasing to a value of 2.0 compared to the values in the run WLM-nodiff. However,  this is probably not driven by the metal diffusion directly, but rather by the increase in mean star formation in the  ``diff'' systems by a factor of about 1.3 to 1.5. Hence, this is an indirect effect of the metal diffusion. We conclude that metal mixing plays little to no role in those quantities discussed above; this is backed up by the mean values for the loading factors at 1 kpc and 10 kpc, which we report in Table \ref{tab:loadings}. The major difference that arises is the phase-space distribution of the metals, as we report in Fig.~\ref{fig:yout}. In those, it turns out that metal diffusion is quite important for obtaining a meaningful phase structure that is not bimodally separated between enriched SN ejecta in the hot and ambient (not metal-enriched) gas in the warm phase of the outflow. More importantly, we find that all the high-metallicity gas located at the peak of the cooling curve is cooling rapidly to about $10^4$ K, causing a dipin the metal distribution phase-space structure. In the runs with metal diffusion, the gap disappears as metals diffuse out of the very hot phase into this intermediate regime, thereby ``filling up'' that gap. In turn, the metals that ``cool off'' from around $10^5$ K are rapidly mixing into the warm outflow, increasing the metal enrichment factor from the ambient value of around 1 in that regime to a value of around 2 when metal diffusion is active. We conclude that metal diffusion models should always be applied whenever the research objective centers on the detailed phase-space distribution of metals.

\subsection{Pathways for improvements}

The problem that we encounter for the metal phase-space distribution is independent of the question of whether or not a metal diffusion model should be adopted  but rather has to do with the neighbor-weighted mass spreading scheme that we use for the mass and energy distribution of each supernova event. Some previous simulations (including our own) that do not include a metal mixing model, specifically at our solar mass resolution scale, were chosen on the grounds that at some point, dynamical mixing of metal mass would be dominant and metal mixing models would not be needed anymore. However, this turns out to be quite untrue, since in most of our simulations we find that multiple enrichment processes in superbubbles  lead to particles with an unrealistically high metal composition of up to around 20 Z$_{\odot}$ (this is especially problematic in LMC). This fact does not change with active metal diffusion as it stems from the injection scheme that was chosen for these kinds of simulations to begin with. Nevertheless, the maximum metallicity found in simulations with metal diffusion is only around 10 Z$_{\odot}$, which is an improvement, but still a very high value. With active metal diffusion, however, particles with such high metallicities are quickly depleted and reduced to much lower values of around 2-3 Z$_{\odot}$. 
We note that \citet{Hu2019} introduced a healpix scheme for the injection. While this scheme takes the local geometry of the injection region into account, it still does not strongly reduce the issue of ``over-enriched particles'' at injection. Additionally, the loading factor reacts only very weakly in such a scheme, as long as the resolution is high enough (as it is in all of our simulations). It is worthwhile noting that in our MFM scheme, the over-enrichment is somewhat less problematic than in SPH, since we operate MFM with a much smaller neighbor number than SPH and thus it  simply becomes less likely that  the same neighbors will be selected over and over again. 

An interesting direction for future simulations would be to spawn feedback particles directly, as is done in higher-resolution simulations of molecular clouds \citep[e.g.,][]{Grudic2021, Grudic2022}. One advantage of a scheme like that is that only the newly spawned particles would carry the metal mass. In that scenario (and at very high resolution) a dynamical mixing argument might be made and the need for metal diffusion models might be obsolete. However, such a scheme must  first be implemented and carefully tested on our resolution scale. 

\section{Conclusions}
\label{sec:conclusions}

In this paper, we carried out a set of dwarf galaxy simulations with and without a sub-grid model for metal diffusion. All of our simulations employ an ISM model that explicitly resolves individual SN explosions and other feedback channels within a multiphase ISM. Our main findings are as follows: 

\begin{enumerate}
    \item We find a slight increase in the mean star formation rate, by a factor of 1.3 to 1.5.
    \item The impact of metal diffusion on the phase structure of the ISM is minor, but we find that there is slightly more metal mass in both the hot and the cold ISM while metal diffusion is active. Subsequently, metals at low temperatures ($\sim$ 100K) and intermediate number densities ($\sim 0.1$ cm$^{-3}$) are removed from the phase space and more efficiently mixed into the cold dense part of the ISM.
    \item There is no effect on the absolute values of the loading factors for mass, metals and energy. 
    \item The largest effect following the inclusion of metal diffusion is in the phase-space structure in the joint outflow velocity-sound speed space. Without metal diffusion, we find a clear phase-space separation of hot, highly enriched gas (y$_z^{out} \approx 3.5-4$) from warm ambient gas (y$_z^{out} \approx$ 1). When metal diffusion is active, the simulations establish a clear gradient and mix the highly enriched material with the ambient gas, leading to a phase-space structure that is more similar to results obtained by Eulerian methods with intrinsic metal diffusion stemming from the Riemann solver.    
\end{enumerate}

Finally, we want to point out the importance of utilizing metal diffusion models when making predictions for outflow rates using realistic absorption line tracers. We are convinced that this can only be done using models based on SPH simulations that directly include a treatment for metal diffusion based on the phase space obtained in Fig.~\ref{fig:metal_enrichement_out}; this is the key finding of this study. If one were to model the absorption line features in our simulations without metal diffusion  and attempt to reconstruct outflow rates based on these absorption line features, one would likely overestimate the resulting outflow rates in the models without metal diffusion; we will test this in more detail in future work. Additionally, we note, that strictly speaking, the importance of metal diffusion models for Lagrangian-methods does not stop at the metal phase-structure. A similar point could be made for metal abundance patterns in the ISM itself and there is evidence for this from the Fire-2 simulations \citep[see e.g.;][for more information on that particular point]{Escala2018}.

\section*{Acknowledgements}
We thank Greg L.~Bryan, Eve C.~Ostriker, R\"udiger Pakmor, Rachel S.~Somerville, Laura Sommovigo and Romain Teyssier for useful discussions. UPS, DR and MEO are supported by the Simons Foundation. We thank Naomi Klarreich for assistance with preparation of the manuscript.

\vspace{5mm}
\facilities{Rusty (Flatiron Institute)}

\software{P-Gadget3 \citep[][]{springel05, Hu2014, Steinwandel2020}, astropy \citep{2013A&A...558A..33A,2018AJ....156..123A}, CMasher \citep{cmasher}, matplotlib \citep{Matplotlib}, scipy \citep[][]{scipy}, numpy \citep[][]{numpy}}

\bibliography{sample631}{}

\begin{thebibliography}{}
\expandafter\ifx\csname natexlab\endcsname\relax\def\natexlab#1{#1}\fi
\providecommand{\url}[1]{\href{#1}{#1}}
\providecommand{\dodoi}[1]{doi:~\href{http://doi.org/#1}{\nolinkurl{#1}}}
\providecommand{\doeprint}[1]{\href{http://ascl.net/#1}{\nolinkurl{http://ascl.net/#1}}}
\providecommand{\doarXiv}[1]{\href{https://arxiv.org/abs/#1}{\nolinkurl{https://arxiv.org/abs/#1}}}

\bibitem[{{Arellano-C{\'o}rdova} {et~al.}(2021){Arellano-C{\'o}rdova}, {Esteban}, {Garc{\'\i}a-Rojas}, \& {M{\'e}ndez-Delgado}}]{Arellano2021}
{Arellano-C{\'o}rdova}, K.~Z., {Esteban}, C., {Garc{\'\i}a-Rojas}, J., \& {M{\'e}ndez-Delgado}, J.~E. 2021, \mnras, 502, 225, \dodoi{10.1093/mnras/staa3903}

\bibitem[{{Astropy Collaboration} {et~al.}(2013){Astropy Collaboration}, {Robitaille}, {Tollerud}, {Greenfield}, {Droettboom}, {Bray}, {Aldcroft}, {Davis}, {Ginsburg}, {Price-Whelan}, {Kerzendorf}, {Conley}, {Crighton}, {Barbary}, {Muna}, {Ferguson}, {Grollier}, {Parikh}, {Nair}, {Unther}, {Deil}, {Woillez}, {Conseil}, {Kramer}, {Turner}, {Singer}, {Fox}, {Weaver}, {Zabalza}, {Edwards}, {Azalee Bostroem}, {Burke}, {Casey}, {Crawford}, {Dencheva}, {Ely}, {Jenness}, {Labrie}, {Lim}, {Pierfederici}, {Pontzen}, {Ptak}, {Refsdal}, {Servillat}, \& {Streicher}}]{2013A&A...558A..33A}
{Astropy Collaboration}, {Robitaille}, T.~P., {Tollerud}, E.~J., {et~al.} 2013, \aap, 558, A33, \dodoi{10.1051/0004-6361/201322068}

\bibitem[{{Astropy Collaboration} {et~al.}(2018){Astropy Collaboration}, {Price-Whelan}, {Sip{\H{o}}cz}, {G{\"u}nther}, {Lim}, {Crawford}, {Conseil}, {Shupe}, {Craig}, {Dencheva}, {Ginsburg}, {VanderPlas}, {Bradley}, {P{\'e}rez-Su{\'a}rez}, {de Val-Borro}, {Aldcroft}, {Cruz}, {Robitaille}, {Tollerud}, {Ardelean}, {Babej}, {Bach}, {Bachetti}, {Bakanov}, {Bamford}, {Barentsen}, {Barmby}, {Baumbach}, {Berry}, {Biscani}, {Boquien}, {Bostroem}, {Bouma}, {Brammer}, {Bray}, {Breytenbach}, {Buddelmeijer}, {Burke}, {Calderone}, {Cano Rodr{\'\i}guez}, {Cara}, {Cardoso}, {Cheedella}, {Copin}, {Corrales}, {Crichton}, {D'Avella}, {Deil}, {Depagne}, {Dietrich}, {Donath}, {Droettboom}, {Earl}, {Erben}, {Fabbro}, {Ferreira}, {Finethy}, {Fox}, {Garrison}, {Gibbons}, {Goldstein}, {Gommers}, {Greco}, {Greenfield}, {Groener}, {Grollier}, {Hagen}, {Hirst}, {Homeier}, {Horton}, {Hosseinzadeh}, {Hu}, {Hunkeler}, {Ivezi{\'c}}, {Jain}, {Jenness}, {Kanarek}, {Kendrew}, {Kern}, {Kerzendorf}, {Khvalko}, {King}, {Kirkby}, {Kulkarni},
  {Kumar}, {Lee}, {Lenz}, {Littlefair}, {Ma}, {Macleod}, {Mastropietro}, {McCully}, {Montagnac}, {Morris}, {Mueller}, {Mumford}, {Muna}, {Murphy}, {Nelson}, {Nguyen}, {Ninan}, {N{\"o}the}, {Ogaz}, {Oh}, {Parejko}, {Parley}, {Pascual}, {Patil}, {Patil}, {Plunkett}, {Prochaska}, {Rastogi}, {Reddy Janga}, {Sabater}, {Sakurikar}, {Seifert}, {Sherbert}, {Sherwood-Taylor}, {Shih}, {Sick}, {Silbiger}, {Singanamalla}, {Singer}, {Sladen}, {Sooley}, {Sornarajah}, {Streicher}, {Teuben}, {Thomas}, {Tremblay}, {Turner}, {Terr{\'o}n}, {van Kerkwijk}, {de la Vega}, {Watkins}, {Weaver}, {Whitmore}, {Woillez}, {Zabalza}, \& {Astropy Contributors}}]{2018AJ....156..123A}
{Astropy Collaboration}, {Price-Whelan}, A.~M., {Sip{\H{o}}cz}, B.~M., {et~al.} 2018, \aj, 156, 123, \dodoi{10.3847/1538-3881/aabc4f}

\bibitem[{{Aumer} \& {White}(2013)}]{Aumer2013}
{Aumer}, M., \& {White}, S.~D.~M. 2013, \mnras, 428, 1055, \dodoi{10.1093/mnras/sts083}

\bibitem[{{Barnes} \& {Hut}(1986)}]{Barnes1986}
{Barnes}, J., \& {Hut}, P. 1986, \nat, 324, 446, \dodoi{10.1038/324446a0}

\bibitem[{{Bourne} \& {Sijacki}(2017)}]{Bourne2017}
{Bourne}, M.~A., \& {Sijacki}, D. 2017, \mnras, 472, 4707, \dodoi{10.1093/mnras/stx2269}

\bibitem[{{Brook} {et~al.}(2014){Brook}, {Stinson}, {Gibson}, {Shen}, {Macci{\`o}}, {Obreja}, {Wadsley}, \& {Quinn}}]{Brook2014}
{Brook}, C.~B., {Stinson}, G., {Gibson}, B.~K., {et~al.} 2014, \mnras, 443, 3809, \dodoi{10.1093/mnras/stu1406}

\bibitem[{{Carr} {et~al.}(2023){Carr}, {Bryan}, {Fielding}, {Pandya}, \& {Somerville}}]{Carr2023}
{Carr}, C., {Bryan}, G.~L., {Fielding}, D.~B., {Pandya}, V., \& {Somerville}, R.~S. 2023, \apj, 949, 21, \dodoi{10.3847/1538-4357/acc4c7}

\bibitem[{{Clark} {et~al.}(2012){Clark}, {Glover}, \& {Klessen}}]{Clark2012}
{Clark}, P.~C., {Glover}, S.~C.~O., \& {Klessen}, R.~S. 2012, \mnras, 420, 745, \dodoi{10.1111/j.1365-2966.2011.20087.x}

\bibitem[{{Colbrook} {et~al.}(2017){Colbrook}, {Ma}, {Hopkins}, \& {Squire}}]{Colbrook2017}
{Colbrook}, M.~J., {Ma}, X., {Hopkins}, P.~F., \& {Squire}, J. 2017, \mnras, 467, 2421, \dodoi{10.1093/mnras/stx261}

\bibitem[{{Dav{\'e}} {et~al.}(2012){Dav{\'e}}, {Finlator}, \& {Oppenheimer}}]{Dave2012}
{Dav{\'e}}, R., {Finlator}, K., \& {Oppenheimer}, B.~D. 2012, \mnras, 421, 98, \dodoi{10.1111/j.1365-2966.2011.20148.x}

\bibitem[{{Draine}(2011)}]{Draine2011}
{Draine}, B.~T. 2011, {Physics of the Interstellar and Intergalactic Medium} (Princeton University Press)

\bibitem[{{Elmegreen} \& {Scalo}(2004)}]{Elmegreen2004}
{Elmegreen}, B.~G., \& {Scalo}, J. 2004, \araa, 42, 211, \dodoi{10.1146/annurev.astro.41.011802.094859}

\bibitem[{{Erkal} {et~al.}(2019){Erkal}, {Belokurov}, {Laporte}, {Koposov}, {Li}, {Grillmair}, {Kallivayalil}, {Price-Whelan}, {Evans}, {Hawkins}, {Hendel}, {Mateu}, {Navarro}, {del Pino}, {Slater}, {Sohn}, \& {Orphan Aspen Treasury Collaboration}}]{Erkal2019}
{Erkal}, D., {Belokurov}, V., {Laporte}, C.~F.~P., {et~al.} 2019, \mnras, 487, 2685, \dodoi{10.1093/mnras/stz1371}

\bibitem[{{Escala} {et~al.}(2018){Escala}, {Wetzel}, {Kirby}, {Hopkins}, {Ma}, {Wheeler}, {Kere{\v{s}}}, {Faucher-Gigu{\`e}re}, \& {Quataert}}]{Escala2018}
{Escala}, I., {Wetzel}, A., {Kirby}, E.~N., {et~al.} 2018, \mnras, 474, 2194, \dodoi{10.1093/mnras/stx2858}

\bibitem[{{Esteban} \& {Garc{\'\i}a-Rojas}(2018)}]{Esteban2018}
{Esteban}, C., \& {Garc{\'\i}a-Rojas}, J. 2018, \mnras, 478, 2315, \dodoi{10.1093/mnras/sty1168}

\bibitem[{{Esteban} {et~al.}(2022){Esteban}, {M{\'e}ndez-Delgado}, {Garc{\'\i}a-Rojas}, \& {Arellano-C{\'o}rdova}}]{Esteban2022}
{Esteban}, C., {M{\'e}ndez-Delgado}, J.~E., {Garc{\'\i}a-Rojas}, J., \& {Arellano-C{\'o}rdova}, K.~Z. 2022, \apj, 931, 92, \dodoi{10.3847/1538-4357/ac6b38}

\bibitem[{{Faucher-Gigu{\`e}re} \& {Oh}(2023)}]{Faucher2023}
{Faucher-Gigu{\`e}re}, C.-A., \& {Oh}, S.~P. 2023, \araa, 61, 131, \dodoi{10.1146/annurev-astro-052920-125203}

\bibitem[{{Fielding} {et~al.}(2018){Fielding}, {Quataert}, \& {Martizzi}}]{Fielding2018}
{Fielding}, D., {Quataert}, E., \& {Martizzi}, D. 2018, \mnras, 481, 3325, \dodoi{10.1093/mnras/sty2466}

\bibitem[{{Fielding} {et~al.}(2017{\natexlab{a}}){Fielding}, {Quataert}, {Martizzi}, \& {Faucher-Gigu{\`e}re}}]{Fielding2017}
{Fielding}, D., {Quataert}, E., {Martizzi}, D., \& {Faucher-Gigu{\`e}re}, C.-A. 2017{\natexlab{a}}, \mnras, 470, L39, \dodoi{10.1093/mnrasl/slx072}

\bibitem[{{Fielding} {et~al.}(2017{\natexlab{b}}){Fielding}, {Quataert}, {McCourt}, \& {Thompson}}]{Fielding2017_cgm}
{Fielding}, D., {Quataert}, E., {McCourt}, M., \& {Thompson}, T.~A. 2017{\natexlab{b}}, \mnras, 466, 3810, \dodoi{10.1093/mnras/stw3326}

\bibitem[{{Fielding} {et~al.}(2020){Fielding}, {Tonnesen}, {DeFelippis}, {Li}, {Su}, {Bryan}, {Kim}, {Forbes}, {Somerville}, {Battaglia}, {Schneider}, {Li}, {Choi}, {Hayward}, \& {Hernquist}}]{Fielding2020}
{Fielding}, D.~B., {Tonnesen}, S., {DeFelippis}, D., {et~al.} 2020, \apj, 903, 32, \dodoi{10.3847/1538-4357/abbc6d}

\bibitem[{{Gaburov} \& {Nitadori}(2011)}]{Gaburov2011}
{Gaburov}, E., \& {Nitadori}, K. 2011, \mnras, 414, 129, \dodoi{10.1111/j.1365-2966.2011.18313.x}

\bibitem[{{Gatto} {et~al.}(2017){Gatto}, {Walch}, {Naab}, {Girichidis}, {W{\"u}nsch}, {Glover}, {Klessen}, {Clark}, {Peters}, {Derigs}, {Baczynski}, \& {Puls}}]{Gatto2017}
{Gatto}, A., {Walch}, S., {Naab}, T., {et~al.} 2017, \mnras, 466, 1903, \dodoi{10.1093/mnras/stw3209}

\bibitem[{{Georgy} {et~al.}(2013){Georgy}, {Ekstr{\"o}m}, {Eggenberger}, {Meynet}, {Haemmerl{\'e}}, {Maeder}, {Granada}, {Groh}, {Hirschi}, {Mowlavi}, {Yusof}, {Charbonnel}, {Decressin}, \& {Barblan}}]{Georgy2013}
{Georgy}, C., {Ekstr{\"o}m}, S., {Eggenberger}, P., {et~al.} 2013, \aap, 558, A103, \dodoi{10.1051/0004-6361/201322178}

\bibitem[{{Girichidis} {et~al.}(2018){Girichidis}, {Seifried}, {Naab}, {Peters}, {Walch}, {W{\"u}nsch}, {Glover}, \& {Klessen}}]{Girichidis2018}
{Girichidis}, P., {Seifried}, D., {Naab}, T., {et~al.} 2018, \mnras, 480, 3511, \dodoi{10.1093/mnras/sty2016}

\bibitem[{{Girichidis} {et~al.}(2016){Girichidis}, {Walch}, {Naab}, {Gatto}, {W{\"u}nsch}, {Glover}, {Klessen}, {Clark}, {Peters}, {Derigs}, \& {Baczynski}}]{Girichidis2016}
{Girichidis}, P., {Walch}, S., {Naab}, T., {et~al.} 2016, \mnras, 456, 3432, \dodoi{10.1093/mnras/stv2742}

\bibitem[{{Glover} \& {Clark}(2012)}]{Glover2012}
{Glover}, S.~C.~O., \& {Clark}, P.~C. 2012, \mnras, 421, 9, \dodoi{10.1111/j.1365-2966.2011.19648.x}

\bibitem[{{Glover} \& {Mac Low}(2007{\natexlab{a}})}]{Glover2007a}
{Glover}, S.~C.~O., \& {Mac Low}, M.-M. 2007{\natexlab{a}}, \apjs, 169, 239, \dodoi{10.1086/512238}

\bibitem[{{Glover} \& {Mac Low}(2007{\natexlab{b}})}]{Glover2007b}
---. 2007{\natexlab{b}}, \apj, 659, 1317, \dodoi{10.1086/512227}

\bibitem[{{Gong} {et~al.}(2017){Gong}, {Ostriker}, \& {Wolfire}}]{Gong2016}
{Gong}, M., {Ostriker}, E.~C., \& {Wolfire}, M.~G. 2017, \apj, 843, 38, \dodoi{10.3847/1538-4357/aa7561}

\bibitem[{{Greif} {et~al.}(2008){Greif}, {Johnson}, {Klessen}, \& {Bromm}}]{Greif2008}
{Greif}, T.~H., {Johnson}, J.~L., {Klessen}, R.~S., \& {Bromm}, V. 2008, \mnras, 387, 1021, \dodoi{10.1111/j.1365-2966.2008.13326.x}

\bibitem[{{Grudi{\'c}} {et~al.}(2021){Grudi{\'c}}, {Guszejnov}, {Hopkins}, {Offner}, \& {Faucher-Gigu{\`e}re}}]{Grudic2021}
{Grudi{\'c}}, M.~Y., {Guszejnov}, D., {Hopkins}, P.~F., {Offner}, S. S.~R., \& {Faucher-Gigu{\`e}re}, C.-A. 2021, \mnras, 506, 2199, \dodoi{10.1093/mnras/stab1347}

\bibitem[{{Grudi{\'c}} {et~al.}(2022){Grudi{\'c}}, {Guszejnov}, {Offner}, {Rosen}, {Raju}, {Faucher-Gigu{\`e}re}, \& {Hopkins}}]{Grudic2022}
{Grudi{\'c}}, M.~Y., {Guszejnov}, D., {Offner}, S. S.~R., {et~al.} 2022, arXiv e-prints, arXiv:2201.00882.
\newblock \doarXiv{2201.00882}

\bibitem[{{Gutcke} {et~al.}(2021){Gutcke}, {Pakmor}, {Naab}, \& {Springel}}]{Gutcke2021}
{Gutcke}, T.~A., {Pakmor}, R., {Naab}, T., \& {Springel}, V. 2021, \mnras, 501, 5597, \dodoi{10.1093/mnras/staa3875}

\bibitem[{{Hafen} {et~al.}(2019){Hafen}, {Faucher-Gigu{\`e}re}, {Angl{\'e}s-Alc{\'a}zar}, {Stern}, {Kere{\v{s}}}, {Hummels}, {Esmerian}, {Garrison-Kimmel}, {El-Badry}, {Wetzel}, {Chan}, {Hopkins}, \& {Murray}}]{Hafen2019}
{Hafen}, Z., {Faucher-Gigu{\`e}re}, C.-A., {Angl{\'e}s-Alc{\'a}zar}, D., {et~al.} 2019, \mnras, 488, 1248, \dodoi{10.1093/mnras/stz1773}

\bibitem[{{Hafen} {et~al.}(2020){Hafen}, {Faucher-Gigu{\`e}re}, {Angl{\'e}s-Alc{\'a}zar}, {Stern}, {Kere{\v{s}}}, {Esmerian}, {Wetzel}, {El-Badry}, {Chan}, \& {Murray}}]{Hafen2020}
---. 2020, \mnras, 494, 3581, \dodoi{10.1093/mnras/staa902}

\bibitem[{Harris {et~al.}(2020)Harris, Millman, van~der Walt, Gommers, Virtanen, Cournapeau, Wieser, Taylor, Berg, Smith, Kern, Picus, Hoyer, van Kerkwijk, Brett, Haldane, del R{\'{i}}o, Wiebe, Peterson, G{\'{e}}rard-Marchant, Sheppard, Reddy, Weckesser, Abbasi, Gohlke, \& Oliphant}]{numpy}
Harris, C.~R., Millman, K.~J., van~der Walt, S.~J., {et~al.} 2020, Nature, 585, 357, \dodoi{10.1038/s41586-020-2649-2}

\bibitem[{{Heckman} {et~al.}(2017){Heckman}, {Borthakur}, {Wild}, {Schiminovich}, \& {Bordoloi}}]{Heckman2017}
{Heckman}, T., {Borthakur}, S., {Wild}, V., {Schiminovich}, D., \& {Bordoloi}, R. 2017, \apj, 846, 151, \dodoi{10.3847/1538-4357/aa80dc}

\bibitem[{{Hislop} {et~al.}(2022){Hislop}, {Naab}, {Steinwandel}, {Lah{\'e}n}, {Irodotou}, {Johansson}, \& {Walch}}]{Hislop2022}
{Hislop}, J.~M., {Naab}, T., {Steinwandel}, U.~P., {et~al.} 2022, \mnras, 509, 5938, \dodoi{10.1093/mnras/stab3347}

\bibitem[{{Hopkins}(2015)}]{Hopkins2015}
{Hopkins}, P.~F. 2015, \mnras, 450, 53, \dodoi{10.1093/mnras/stv195}

\bibitem[{{Hopkins} {et~al.}(2022{\natexlab{a}}){Hopkins}, {Butsky}, {Panopoulou}, {Ji}, {Quataert}, {Faucher-Gigu{\`e}re}, \& {Kere{\v{s}}}}]{Hopkins2022_spectra}
{Hopkins}, P.~F., {Butsky}, I.~S., {Panopoulou}, G.~V., {et~al.} 2022{\natexlab{a}}, \mnras, 516, 3470, \dodoi{10.1093/mnras/stac1791}

\bibitem[{{Hopkins} {et~al.}(2021){Hopkins}, {Squire}, {Chan}, {Quataert}, {Ji}, {Kere{\v{s}}}, \& {Faucher-Gigu{\`e}re}}]{Hopkins2021_extrinsic}
{Hopkins}, P.~F., {Squire}, J., {Chan}, T.~K., {et~al.} 2021, \mnras, 501, 4184, \dodoi{10.1093/mnras/staa3691}

\bibitem[{{Hopkins} {et~al.}(2018){Hopkins}, {Wetzel}, {Kere{\v s}}, {Faucher-Gigu{\`e}re}, {Quataert}, {Boylan-Kolchin}, {Murray}, {Hayward}, {Garrison-Kimmel}, {Hummels}, {Feldmann}, {Torrey}, {Ma}, {Angl{\'e}s-Alc{\'a}zar}, {Su}, {Orr}, {Schmitz}, {Escala}, {Sanderson}, {Grudi{\'c}}, {Hafen}, {Kim}, {Fitts}, {Bullock}, {Wheeler}, {Chan}, {Elbert}, \& {Narayanan}}]{Hopkins2018}
{Hopkins}, P.~F., {Wetzel}, A., {Kere{\v s}}, D., {et~al.} 2018, \mnras, 480, 800, \dodoi{10.1093/mnras/sty1690}

\bibitem[{{Hopkins} {et~al.}(2022{\natexlab{b}}){Hopkins}, {Wetzel}, {Wheeler}, {Sanderson}, {Grudic}, {Sameie}, {Boylan-Kolchin}, {Orr}, {Ma}, {Faucher-Giguere}, {Keres}, {Quataert}, {Su}, {Moreno}, {Feldmann}, {Bullock}, {Loebman}, {Angles-Alcazar}, {Stern}, {Necib}, \& {Hayward}}]{Hopkins_fire3}
{Hopkins}, P.~F., {Wetzel}, A., {Wheeler}, C., {et~al.} 2022{\natexlab{b}}, arXiv e-prints, arXiv:2203.00040.
\newblock \doarXiv{2203.00040}

\bibitem[{{Hu}(2019)}]{Hu2019}
{Hu}, C.-Y. 2019, \mnras, 483, 3363, \dodoi{10.1093/mnras/sty3252}

\bibitem[{{Hu} \& {Chiang}(2020)}]{Hu2020}
{Hu}, C.-Y., \& {Chiang}, C.-T. 2020, \apj, 900, 29, \dodoi{10.3847/1538-4357/aba2d5}

\bibitem[{{Hu} {et~al.}(2017){Hu}, {Naab}, {Glover}, {Walch}, \& {Clark}}]{Hu2017}
{Hu}, C.-Y., {Naab}, T., {Glover}, S.~C.~O., {Walch}, S., \& {Clark}, P.~C. 2017, \mnras, 471, 2151, \dodoi{10.1093/mnras/stx1773}

\bibitem[{{Hu} {et~al.}(2016){Hu}, {Naab}, {Walch}, {Glover}, \& {Clark}}]{Hu2016}
{Hu}, C.-Y., {Naab}, T., {Walch}, S., {Glover}, S.~C.~O., \& {Clark}, P.~C. 2016, \mnras, 458, 3528, \dodoi{10.1093/mnras/stw544}

\bibitem[{{Hu} {et~al.}(2014){Hu}, {Naab}, {Walch}, {Moster}, \& {Oser}}]{Hu2014}
{Hu}, C.-Y., {Naab}, T., {Walch}, S., {Moster}, B.~P., \& {Oser}, L. 2014, ArXiv e-prints.
\newblock \doarXiv{1402.1788}

\bibitem[{{Hu} {et~al.}(2022){Hu}, {Smith}, {Teyssier}, {Bryan}, {Verbeke}, {Emerick}, {Somerville}, {Burkhart}, {Li}, {Forbes}, \& {Starkenburg}}]{Hu2023}
{Hu}, C.-Y., {Smith}, M.~C., {Teyssier}, R., {et~al.} 2022, arXiv e-prints, arXiv:2208.10528.
\newblock \doarXiv{2208.10528}

\bibitem[{Hunter(2007)}]{Matplotlib}
Hunter, J.~D. 2007, Computing in Science \& Engineering, 9, 90, \dodoi{10.1109/MCSE.2007.55}

\bibitem[{{Katz} {et~al.}(2022){Katz}, {Liu}, {Kimm}, {Rey}, {Andersson}, {Cameron}, {Rodriguez-Montero}, {Agertz}, {Devriendt}, \& {Slyz}}]{Katz2022}
{Katz}, H., {Liu}, S., {Kimm}, T., {et~al.} 2022, arXiv e-prints, arXiv:2211.04626, \dodoi{10.48550/arXiv.2211.04626}

\bibitem[{{Kim} {et~al.}(2022){Kim}, {Kim}, {Gong}, \& {Ostriker}}]{Kim2022}
{Kim}, C.-G., {Kim}, J.-G., {Gong}, M., \& {Ostriker}, E.~C. 2022, arXiv e-prints, arXiv:2211.13293.
\newblock \doarXiv{2211.13293}

\bibitem[{{Kim} \& {Ostriker}(2015)}]{Kim2015_outflows}
{Kim}, C.-G., \& {Ostriker}, E.~C. 2015, \apj, 815, 67, \dodoi{10.1088/0004-637X/815/1/67}

\bibitem[{{Kim} {et~al.}(2020){Kim}, {Ostriker}, {Fielding}, {Smith}, {Bryan}, {Somerville}, {Forbes}, {Genel}, \& {Hernquist}}]{Kim2020}
{Kim}, C.-G., {Ostriker}, E.~C., {Fielding}, D.~B., {et~al.} 2020, \apjl, 903, L34, \dodoi{10.3847/2041-8213/abc252}

\bibitem[{{Kim} {et~al.}(2023){Kim}, {Gong}, {Kim}, \& {Ostriker}}]{Kim2023}
{Kim}, J.-G., {Gong}, M., {Kim}, C.-G., \& {Ostriker}, E.~C. 2023, \apjs, 264, 10, \dodoi{10.3847/1538-4365/ac9b1d}

\bibitem[{{Kim} {et~al.}(2014){Kim}, {Abel}, {Agertz}, {Bryan}, {Ceverino}, {Christensen}, {Conroy}, {Dekel}, {Gnedin}, {Goldbaum}, {Guedes}, {Hahn}, {Hobbs}, {Hopkins}, {Hummels}, {Iannuzzi}, {Keres}, {Klypin}, {Kravtsov}, {Krumholz}, {Kuhlen}, {Leitner}, {Madau}, {Mayer}, {Moody}, {Nagamine}, {Norman}, {Onorbe}, {O'Shea}, {Pillepich}, {Primack}, {Quinn}, {Read}, {Robertson}, {Rocha}, {Rudd}, {Shen}, {Smith}, {Szalay}, {Teyssier}, {Thompson}, {Todoroki}, {Turk}, {Wadsley}, {Wise}, {Zolotov}, \& {AGORA Collaboration29}}]{Kim2014_agora}
{Kim}, J.-h., {Abel}, T., {Agertz}, O., {et~al.} 2014, \apjs, 210, 14, \dodoi{10.1088/0067-0049/210/1/14}

\bibitem[{{Kim} {et~al.}(2016){Kim}, {Agertz}, {Teyssier}, {Butler}, {Ceverino}, {Choi}, {Feldmann}, {Keller}, {Lupi}, {Quinn}, {Revaz}, {Wallace}, {Gnedin}, {Leitner}, {Shen}, {Smith}, {Thompson}, {Turk}, {Abel}, {Arraki}, {Benincasa}, {Chakrabarti}, {DeGraf}, {Dekel}, {Goldbaum}, {Hopkins}, {Hummels}, {Klypin}, {Li}, {Madau}, {Mandelker}, {Mayer}, {Nagamine}, {Nickerson}, {O'Shea}, {Primack}, {Roca-F{\`a}brega}, {Semenov}, {Shimizu}, {Simpson}, {Todoroki}, {Wadsley}, {Wise}, \& {AGORA Collaboration}}]{Kim2016_agora}
{Kim}, J.-h., {Agertz}, O., {Teyssier}, R., {et~al.} 2016, \apj, 833, 202, \dodoi{10.3847/1538-4357/833/2/202}

\bibitem[{{Lah{\'e}n} {et~al.}(2019{\natexlab{a}}){Lah{\'e}n}, {Naab}, {Johansson}, {Elmegreen}, {Hu}, \& {Walch}}]{Lahen2020a}
{Lah{\'e}n}, N., {Naab}, T., {Johansson}, P.~H., {et~al.} 2019{\natexlab{a}}, \apjl, 879, L18, \dodoi{10.3847/2041-8213/ab2a13}

\bibitem[{{Lah{\'e}n} {et~al.}(2019{\natexlab{b}}){Lah{\'e}n}, {Naab}, {Johansson}, {Elmegreen}, {Hu}, {Walch}, {Steinwand el}, \& {Moster}}]{Lahen2020b}
---. 2019{\natexlab{b}}, arXiv e-prints, arXiv:1911.05093.
\newblock \doarXiv{1911.05093}

\bibitem[{{Lah{\'e}n} {et~al.}(2022){Lah{\'e}n}, {Naab}, {Kauffmann}, {Sz{\'e}csi}, {Hislop}, {Rantala}, {Kozyreva}, {Walch}, \& {Hu}}]{Lahen2022}
{Lah{\'e}n}, N., {Naab}, T., {Kauffmann}, G., {et~al.} 2022, arXiv e-prints, arXiv:2211.15705.
\newblock \doarXiv{2211.15705}

\bibitem[{{Lah{\'e}n} {et~al.}(2023){Lah{\'e}n}, {Naab}, {Kauffmann}, {Sz{\'e}csi}, {Hislop}, {Rantala}, {Kozyreva}, {Walch}, \& {Hu}}]{Lahen2023}
---. 2023, \mnras, 522, 3092, \dodoi{10.1093/mnras/stad1147}

\bibitem[{{Li} {et~al.}(2017){Li}, {Bryan}, \& {Ostriker}}]{Li2017}
{Li}, M., {Bryan}, G.~L., \& {Ostriker}, J.~P. 2017, \apj, 841, 101, \dodoi{10.3847/1538-4357/aa7263}

\bibitem[{{Li} {et~al.}(2020){Li}, {Li}, {Bryan}, {Ostriker}, \& {Quataert}}]{Li2020}
{Li}, M., {Li}, Y., {Bryan}, G.~L., {Ostriker}, E.~C., \& {Quataert}, E. 2020, \apj, 898, 23, \dodoi{10.3847/1538-4357/ab9c22}

\bibitem[{{Naab} \& {Ostriker}(2017)}]{Naab2017}
{Naab}, T., \& {Ostriker}, J.~P. 2017, \araa, 55, 59, \dodoi{10.1146/annurev-astro-081913-040019}

\bibitem[{{Nelson} \& {Langer}(1997)}]{Nelson1997}
{Nelson}, R.~P., \& {Langer}, W.~D. 1997, \apj, 482, 796, \dodoi{10.1086/304167}

\bibitem[{{Pandya} {et~al.}(2023){Pandya}, {Fielding}, {Bryan}, {Carr}, {Somerville}, {Stern}, {Faucher-Gigu{\`e}re}, {Hafen}, {Angl{\'e}s-Alc{\'a}zar}, \& {Forbes}}]{Pandya2023}
{Pandya}, V., {Fielding}, D.~B., {Bryan}, G.~L., {et~al.} 2023, \apj, 956, 118, \dodoi{10.3847/1538-4357/acf3ea}

\bibitem[{{Peters} {et~al.}(2017){Peters}, {Naab}, {Walch}, {Glover}, {Girichidis}, {Pellegrini}, {Klessen}, {W{\"u}nsch}, {Gatto}, \& {Baczynski}}]{Peters2017}
{Peters}, T., {Naab}, T., {Walch}, S., {et~al.} 2017, \mnras, 466, 3293, \dodoi{10.1093/mnras/stw3216}

\bibitem[{Piomelli \& Liu(1995)}]{Piomelli1994}
Piomelli, U., \& Liu, J. 1995, Physics of fluids, 7, 839

\bibitem[{Pope(2000)}]{Pope2000}
Pope, S.~B. 2000, Turbulent Flows (Cambridge University Press)

\bibitem[{{Prasad} {et~al.}(2015){Prasad}, {Sharma}, \& {Babul}}]{Prasad2015}
{Prasad}, D., {Sharma}, P., \& {Babul}, A. 2015, \apj, 811, 108, \dodoi{10.1088/0004-637X/811/2/108}

\bibitem[{{Prasad} {et~al.}(2018){Prasad}, {Sharma}, \& {Babul}}]{Prasad2018}
---. 2018, \apj, 863, 62, \dodoi{10.3847/1538-4357/aacce8}

\bibitem[{{Prochaska} {et~al.}(2017){Prochaska}, {Werk}, {Worseck}, {Tripp}, {Tumlinson}, {Burchett}, {Fox}, {Fumagalli}, {Lehner}, {Peeples}, \& {Tejos}}]{Prochaska2017}
{Prochaska}, J.~X., {Werk}, J.~K., {Worseck}, G., {et~al.} 2017, \apj, 837, 169, \dodoi{10.3847/1538-4357/aa6007}

\bibitem[{{Rennehan}(2021)}]{Rennehan2021}
{Rennehan}, D. 2021, \mnras, 506, 2836, \dodoi{10.1093/mnras/stab1813}

\bibitem[{{Rennehan} {et~al.}(2019){Rennehan}, {Babul}, {Hopkins}, {Dav{\'e}}, \& {Moa}}]{Rennehan2019}
{Rennehan}, D., {Babul}, A., {Hopkins}, P.~F., {Dav{\'e}}, R., \& {Moa}, B. 2019, \mnras, 483, 3810, \dodoi{10.1093/mnras/sty3376}

\bibitem[{{Rey} {et~al.}(2024){Rey}, {Katz}, {Cameron}, {Devriendt}, \& {Slyz}}]{Rey2024}
{Rey}, M.~P., {Katz}, H.~B., {Cameron}, A.~J., {Devriendt}, J., \& {Slyz}, A. 2024, \mnras, 528, 5412, \dodoi{10.1093/mnras/stae388}

\bibitem[{{Schmidt}(2015)}]{Schmidt2015}
{Schmidt}, W. 2015, Living Reviews in Computational Astrophysics, 1, 2, \dodoi{10.1007/lrca-2015-2}

\bibitem[{{Shen} {et~al.}(2013){Shen}, {Madau}, {Guedes}, {Mayer}, {Prochaska}, \& {Wadsley}}]{Shen2013}
{Shen}, S., {Madau}, P., {Guedes}, J., {et~al.} 2013, \apj, 765, 89, \dodoi{10.1088/0004-637X/765/2/89}

\bibitem[{{Shen} {et~al.}(2010){Shen}, {Wadsley}, \& {Stinson}}]{Shen2010}
{Shen}, S., {Wadsley}, J., \& {Stinson}, G. 2010, \mnras, 407, 1581, \dodoi{10.1111/j.1365-2966.2010.17047.x}

\bibitem[{{Shin} {et~al.}(2021){Shin}, {Kim}, \& {Oh}}]{Shin2021}
{Shin}, E.-J., {Kim}, J.-H., \& {Oh}, B.~K. 2021, \apj, 917, 12, \dodoi{10.3847/1538-4357/abffd0}

\bibitem[{{Smagorinsky}(1963)}]{Smagorinsky1963}
{Smagorinsky}, J. 1963, Monthly Weather Review, 91, 99, \dodoi{10.1175/1520-0493(1963)091<0099:GCEWTP>2.3.CO;2}

\bibitem[{{Smith} {et~al.}(2021){Smith}, {Bryan}, {Somerville}, {Hu}, {Teyssier}, {Burkhart}, \& {Hernquist}}]{Smith2021}
{Smith}, M.~C., {Bryan}, G.~L., {Somerville}, R.~S., {et~al.} 2021, \mnras, 506, 3882, \dodoi{10.1093/mnras/stab1896}

\bibitem[{{Smith} {et~al.}(2024){Smith}, {Fielding}, {Bryan}, {Kim}, {Ostriker}, {Somerville}, {Stern}, {Su}, {Weinberger}, {Hu}, {Forbes}, {Hernquist}, {Burkhart}, \& {Li}}]{Smith2024}
{Smith}, M.~C., {Fielding}, D.~B., {Bryan}, G.~L., {et~al.} 2024, \mnras, 527, 1216, \dodoi{10.1093/mnras/stad3168}

\bibitem[{{Soko{\l}owska} {et~al.}(2018){Soko{\l}owska}, {Babul}, {Mayer}, {Shen}, \& {Madau}}]{Sokolowska2018}
{Soko{\l}owska}, A., {Babul}, A., {Mayer}, L., {Shen}, S., \& {Madau}, P. 2018, \apj, 867, 73, \dodoi{10.3847/1538-4357/aae43a}

\bibitem[{{Somerville} \& {Dav{\'e}}(2015)}]{Somerville2015}
{Somerville}, R.~S., \& {Dav{\'e}}, R. 2015, \araa, 53, 51, \dodoi{10.1146/annurev-astro-082812-140951}

\bibitem[{{Springel}(2005)}]{springel05}
{Springel}, V. 2005, \mnras, 364, 1105, \dodoi{10.1111/j.1365-2966.2005.09655.x}

\bibitem[{{Springel} {et~al.}(2005){Springel}, {Di Matteo}, \& {Hernquist}}]{springel05b}
{Springel}, V., {Di Matteo}, T., \& {Hernquist}, L. 2005, \mnras, 361, 776, \dodoi{10.1111/j.1365-2966.2005.09238.x}

\bibitem[{{Steinwandel} {et~al.}(2022{\natexlab{a}}){Steinwandel}, {Bryan}, {Somerville}, {Hayward}, \& {Burkhart}}]{Steinwandel2022}
{Steinwandel}, U.~P., {Bryan}, G.~L., {Somerville}, R.~S., {Hayward}, C.~C., \& {Burkhart}, B. 2022{\natexlab{a}}, arXiv e-prints, arXiv:2205.09774.
\newblock \doarXiv{2205.09774}

\bibitem[{{Steinwandel} \& {Goldberg}(2023)}]{Steinwandel_Goldberg}
{Steinwandel}, U.~P., \& {Goldberg}, J.~A. 2023, arXiv e-prints, arXiv:2310.11495, \dodoi{10.48550/arXiv.2310.11495}

\bibitem[{{Steinwandel} {et~al.}(2022{\natexlab{b}}){Steinwandel}, {Kim}, {Bryan}, {Ostriker}, {Somerville}, \& {Fielding}}]{SteinwandelLMC}
{Steinwandel}, U.~P., {Kim}, C.-G., {Bryan}, G.~L., {et~al.} 2022{\natexlab{b}}, arXiv e-prints, arXiv:2212.03898, \dodoi{10.48550/arXiv.2212.03898}

\bibitem[{{Steinwandel} {et~al.}(2020){Steinwandel}, {Moster}, {Naab}, {Hu}, \& {Walch}}]{Steinwandel2020}
{Steinwandel}, U.~P., {Moster}, B.~P., {Naab}, T., {Hu}, C.-Y., \& {Walch}, S. 2020, \mnras, \dodoi{10.1093/mnras/staa821}

\bibitem[{{Su} {et~al.}(2017){Su}, {Hopkins}, {Hayward}, {Faucher-Gigu{\`e}re}, {Kere{\v{s}}}, {Ma}, \& {Robles}}]{Su2017}
{Su}, K.-Y., {Hopkins}, P.~F., {Hayward}, C.~C., {et~al.} 2017, \mnras, 471, 144, \dodoi{10.1093/mnras/stx1463}

\bibitem[{{Sukhbold} {et~al.}(2016){Sukhbold}, {Ertl}, {Woosley}, {Brown}, \& {Janka}}]{Sukhbold2016}
{Sukhbold}, T., {Ertl}, T., {Woosley}, S.~E., {Brown}, J.~M., \& {Janka}, H.~T. 2016, \apj, 821, 38, \dodoi{10.3847/0004-637X/821/1/38}

\bibitem[{{Tumlinson} {et~al.}(2017){Tumlinson}, {Peeples}, \& {Werk}}]{Tumlinson2017}
{Tumlinson}, J., {Peeples}, M.~S., \& {Werk}, J.~K. 2017, \araa, 55, 389, \dodoi{10.1146/annurev-astro-091916-055240}

\bibitem[{{Tumlinson} {et~al.}(2011){Tumlinson}, {Thom}, {Werk}, {Prochaska}, {Tripp}, {Weinberg}, {Peeples}, {O'Meara}, {Oppenheimer}, {Meiring}, {Katz}, {Dav{\'e}}, {Ford}, \& {Sembach}}]{Tumlinson2011}
{Tumlinson}, J., {Thom}, C., {Werk}, J.~K., {et~al.} 2011, Science, 334, 948, \dodoi{10.1126/science.1209840}

\bibitem[{{van der Velden}(2020)}]{cmasher}
{van der Velden}, E. 2020, The Journal of Open Source Software, 5, 2004, \dodoi{10.21105/joss.02004}

\bibitem[{Virtanen {et~al.}(2020)Virtanen, Gommers, Oliphant, Haberland, Reddy, Cournapeau, Burovski, Peterson, Weckesser, Bright, {van der Walt}, Brett, Wilson, Millman, Mayorov, Nelson, Jones, Kern, Larson, Carey, Polat, Feng, Moore, {VanderPlas}, Laxalde, Perktold, Cimrman, Henriksen, Quintero, Harris, Archibald, Ribeiro, Pedregosa, {van Mulbregt}, \& {SciPy 1.0 Contributors}}]{scipy}
Virtanen, P., Gommers, R., Oliphant, T.~E., {et~al.} 2020, Nature Methods, 17, 261, \dodoi{10.1038/s41592-019-0686-2}

\bibitem[{{Wadsley} {et~al.}(2008){Wadsley}, {Veeravalli}, \& {Couchman}}]{Wadsley2008}
{Wadsley}, J.~W., {Veeravalli}, G., \& {Couchman}, H.~M.~P. 2008, \mnras, 387, 427, \dodoi{10.1111/j.1365-2966.2008.13260.x}

\bibitem[{{Walch} {et~al.}(2015){Walch}, {Girichidis}, {Naab}, {Gatto}, {Glover}, {W{\"u}nsch}, {Klessen}, {Clark}, {Peters}, {Derigs}, \& {Baczynski}}]{Walch2015}
{Walch}, S., {Girichidis}, P., {Naab}, T., {et~al.} 2015, \mnras, 454, 238, \dodoi{10.1093/mnras/stv1975}

\bibitem[{{Werk} {et~al.}(2016){Werk}, {Prochaska}, {Cantalupo}, {Fox}, {Oppenheimer}, {Tumlinson}, {Tripp}, {Lehner}, \& {McQuinn}}]{Werk2016}
{Werk}, J.~K., {Prochaska}, J.~X., {Cantalupo}, S., {et~al.} 2016, \apj, 833, 54, \dodoi{10.3847/1538-4357/833/1/54}

\bibitem[{{Wiersma} {et~al.}(2009){Wiersma}, {Schaye}, \& {Smith}}]{Wiersma2009}
{Wiersma}, R.~P.~C., {Schaye}, J., \& {Smith}, B.~D. 2009, \mnras, 393, 99, \dodoi{10.1111/j.1365-2966.2008.14191.x}

\bibitem[{{Williamson} {et~al.}(2016){Williamson}, {Martel}, \& {Kawata}}]{Williamson2016}
{Williamson}, D., {Martel}, H., \& {Kawata}, D. 2016, \apj, 822, 91, \dodoi{10.3847/0004-637X/822/2/91}

\bibitem[{{Yang} \& {Reynolds}(2016)}]{Karen2016}
{Yang}, H. Y.~K., \& {Reynolds}, C.~S. 2016, \apj, 829, 90, \dodoi{10.3847/0004-637X/829/2/90}

\bibitem[{{Zahedy} {et~al.}(2021){Zahedy}, {Chen}, {Cooper}, {Boettcher}, {Johnson}, {Rudie}, {Chen}, {Cantalupo}, {Cooksey}, {Faucher-Gigu{\`e}re}, {Greene}, {Lopez}, {Mulchaey}, {Penton}, {Petitjean}, {Putman}, {Rafelski}, {Rauch}, {Schaye}, {Simcoe}, \& {Walth}}]{Zahedy2021}
{Zahedy}, F.~S., {Chen}, H.-W., {Cooper}, T.~M., {et~al.} 2021, \mnras, 506, 877, \dodoi{10.1093/mnras/stab1661}

\end{thebibliography}
\bibliographystyle{aasjournal}

\begin{figure*}
    \centering
    \includegraphics[width=0.49\textwidth]{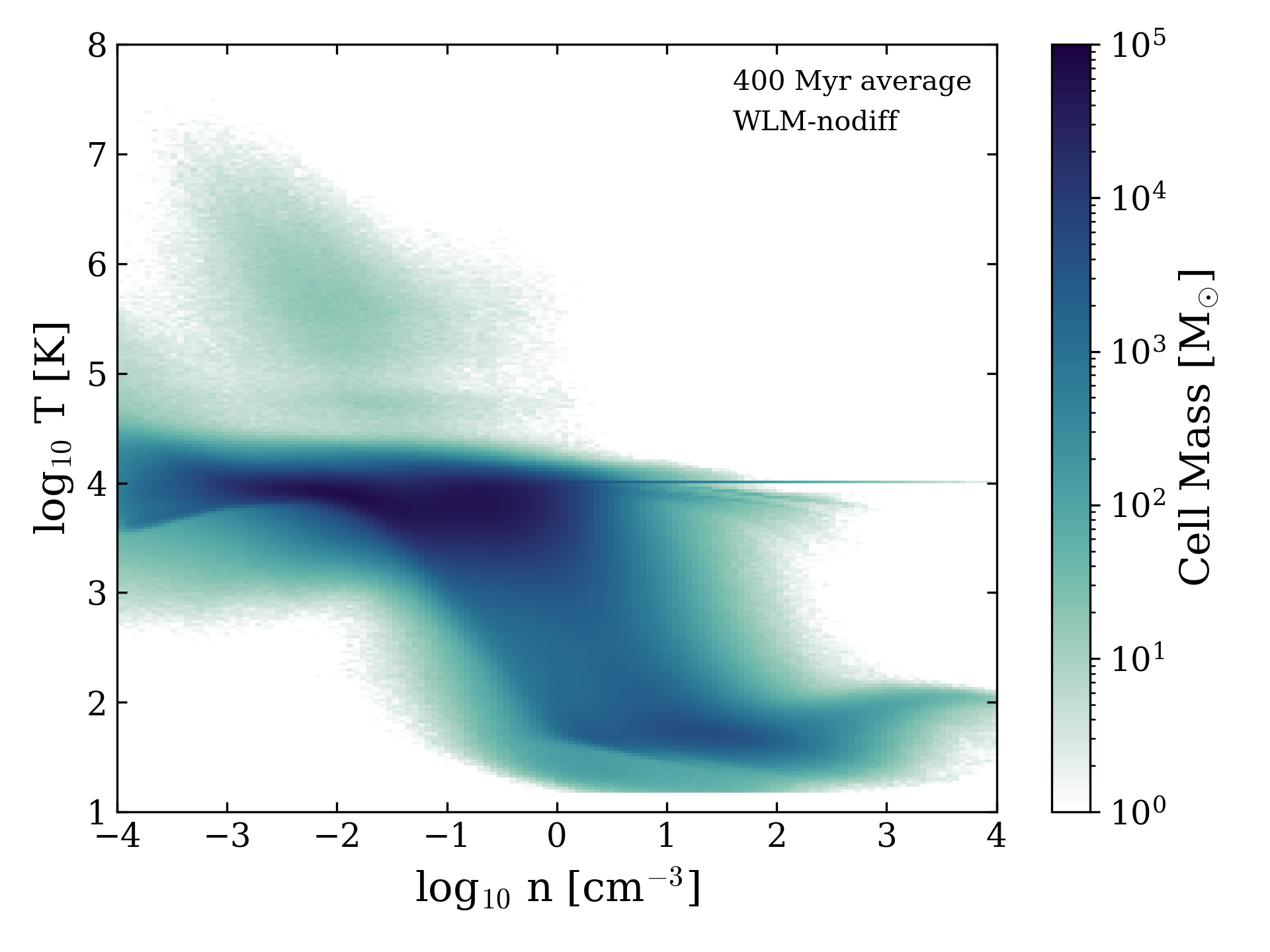}
    \includegraphics[width=0.49\textwidth]{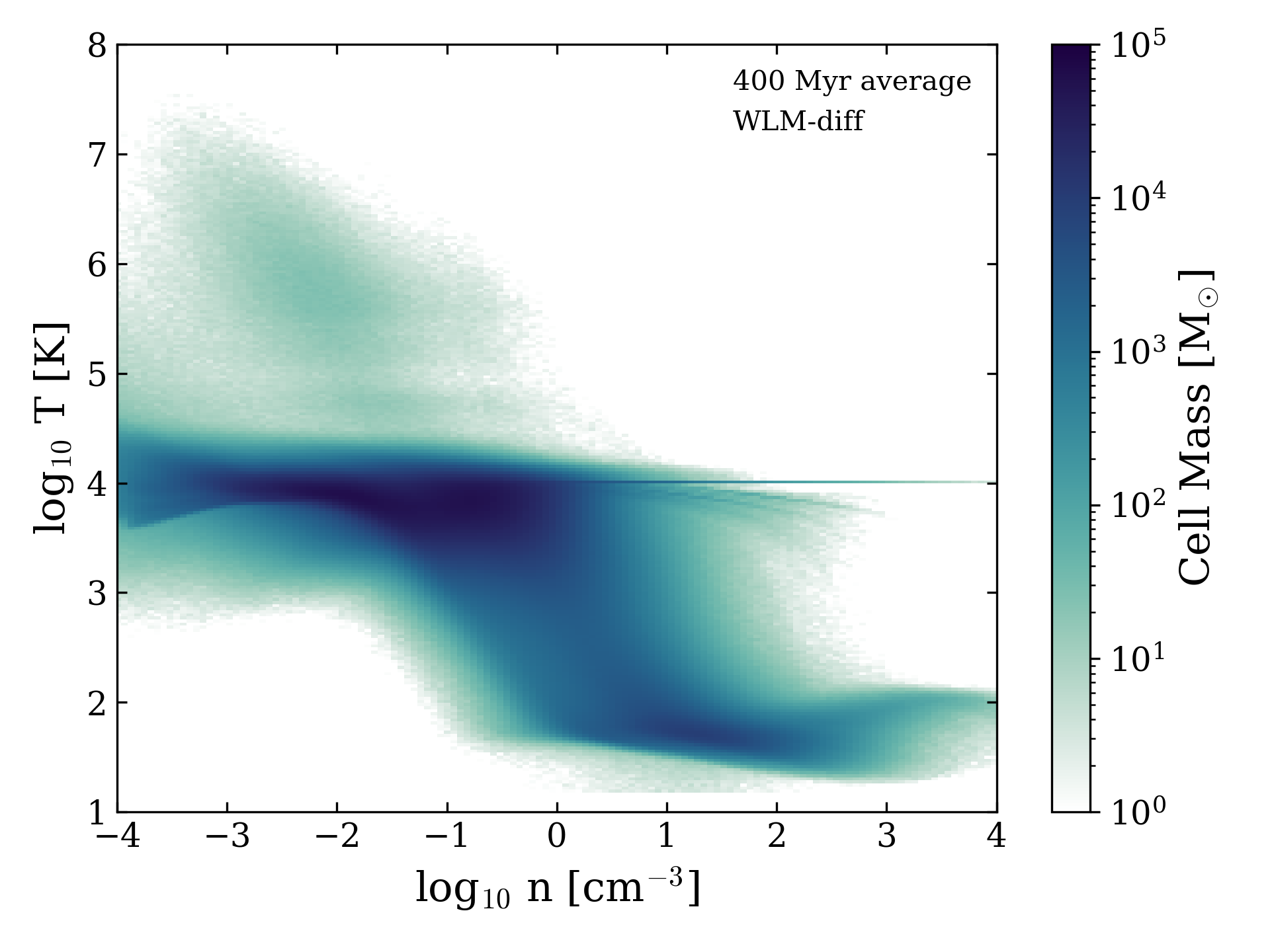}
    \caption{Density-temperature phase-space diagrams, color coded by total mass for the model WLM-nodiff (left) and the model WLM-diff (right).}
    \label{fig:phase_temp}
\end{figure*}

\begin{figure*}
    \centering
    \includegraphics[width=0.49\textwidth]{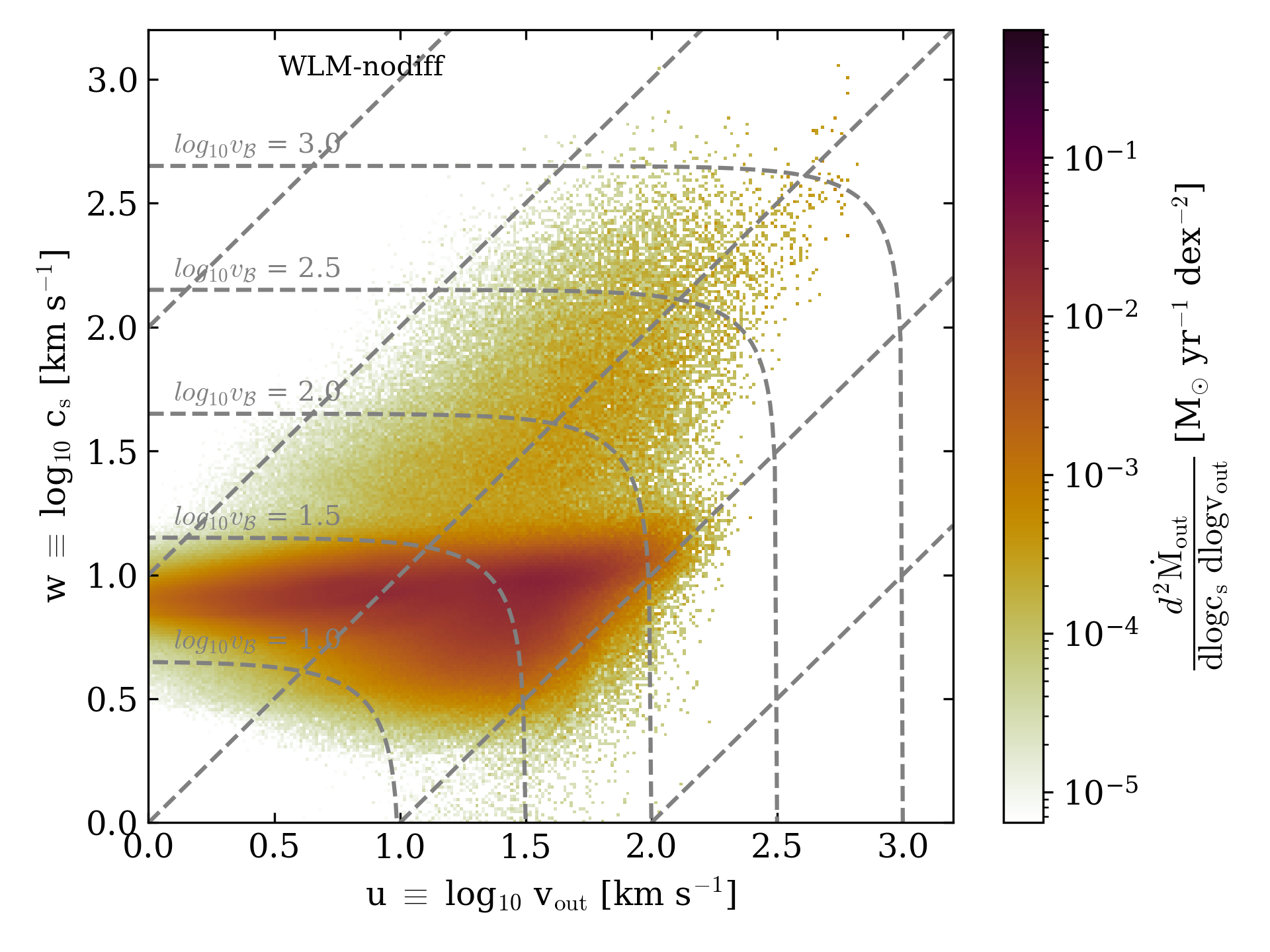}
    \includegraphics[width=0.49\textwidth]{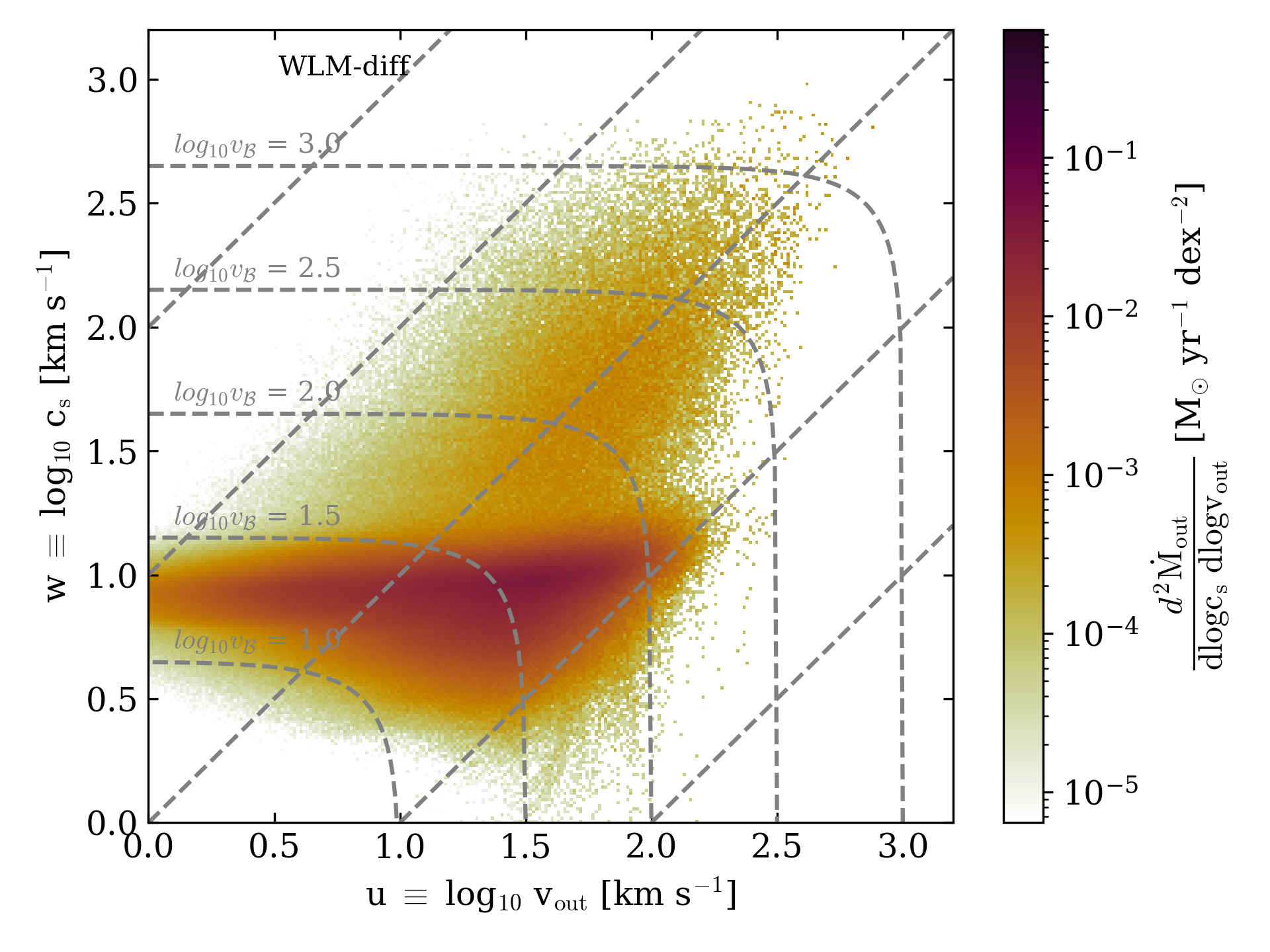}
    \includegraphics[width=0.49\textwidth]{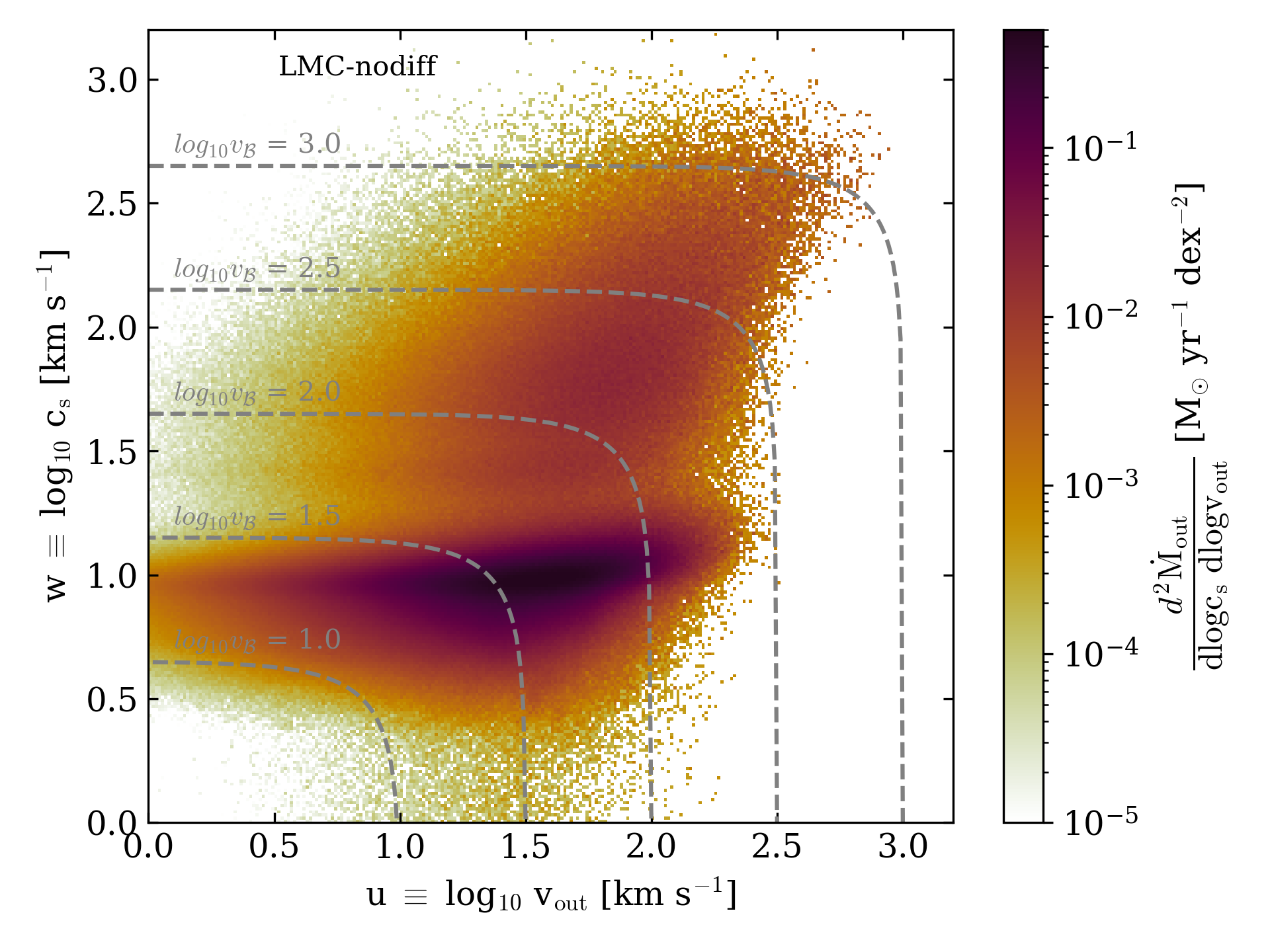}
    \includegraphics[width=0.49\textwidth]{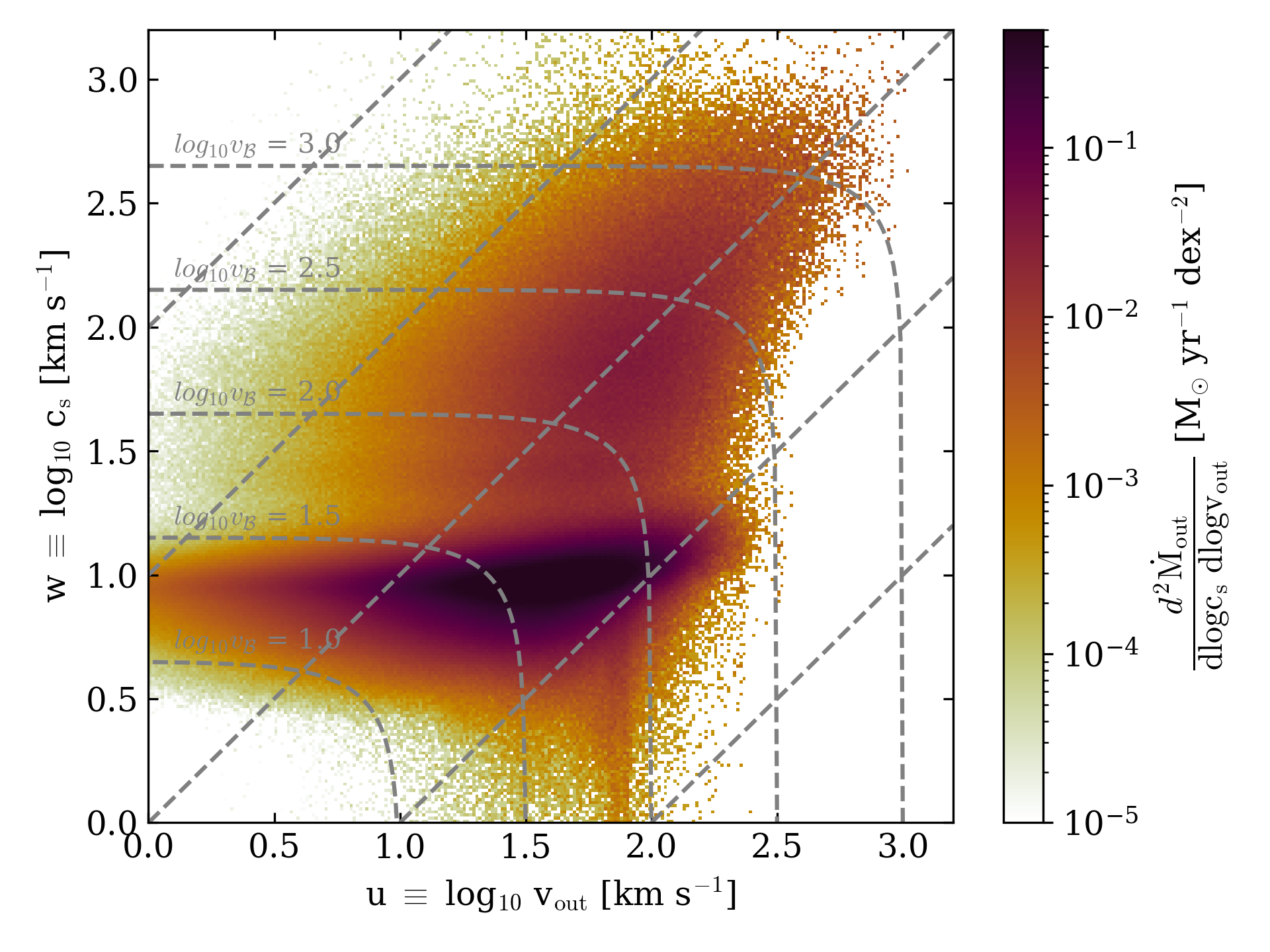}
    \caption{Joint 2D PDFs of outflow velocity and sound speed, color coded by the mass outflow rate for the models WLM-nodiff (top left), WLM-diff (top right), LMC-nodiff (bottom left) and LMC-diff (bottom right). We find that the sound speed in the hot phase of the wind is higher by a factor of two when metal diffusion is enabled. We note that the color bar is on a linear scale.}
    \label{fig:mass_out}
\end{figure*}

\begin{figure*}
    \centering
    \includegraphics[width=0.49\textwidth]{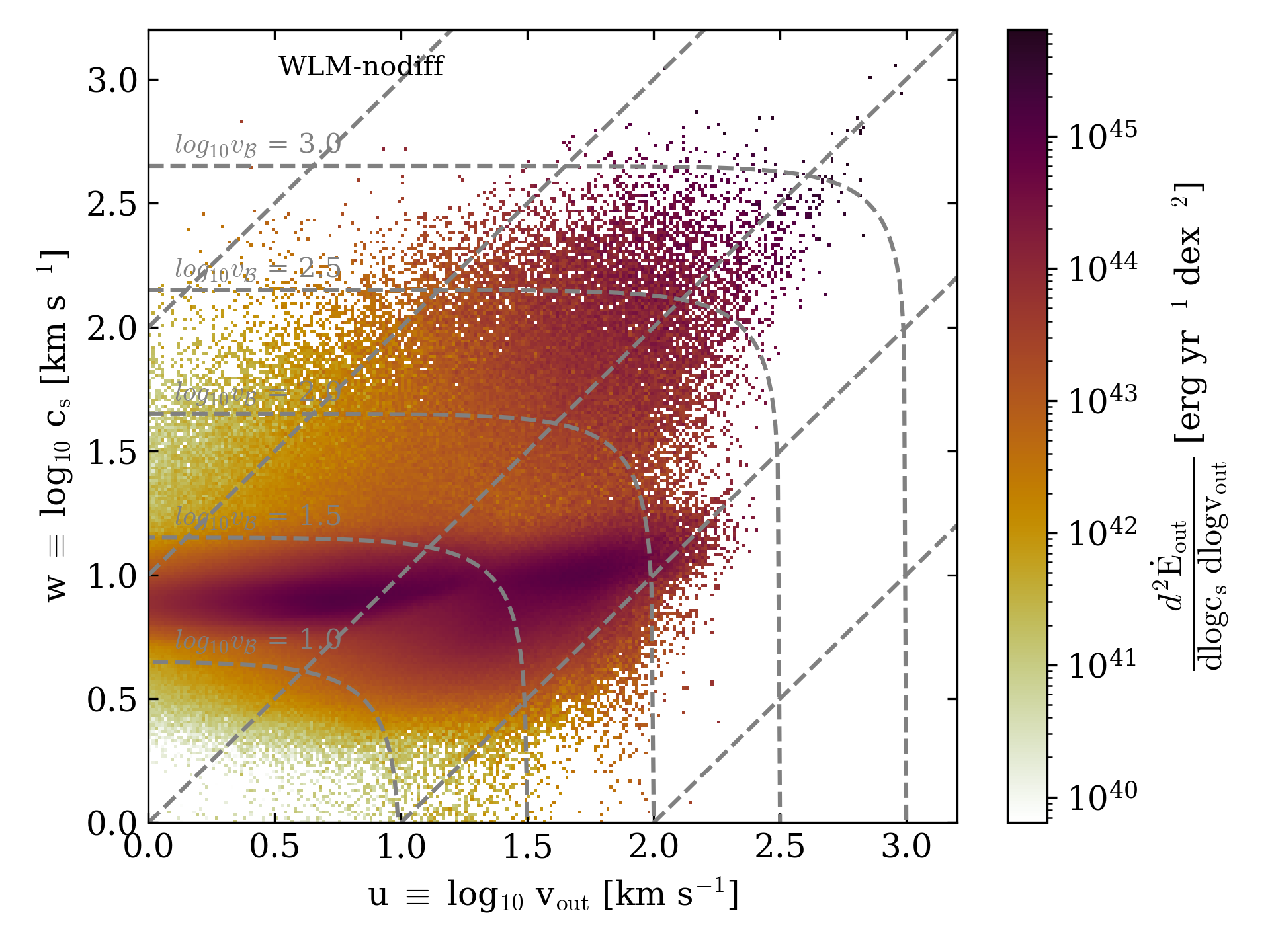}
    \includegraphics[width=0.49\textwidth]{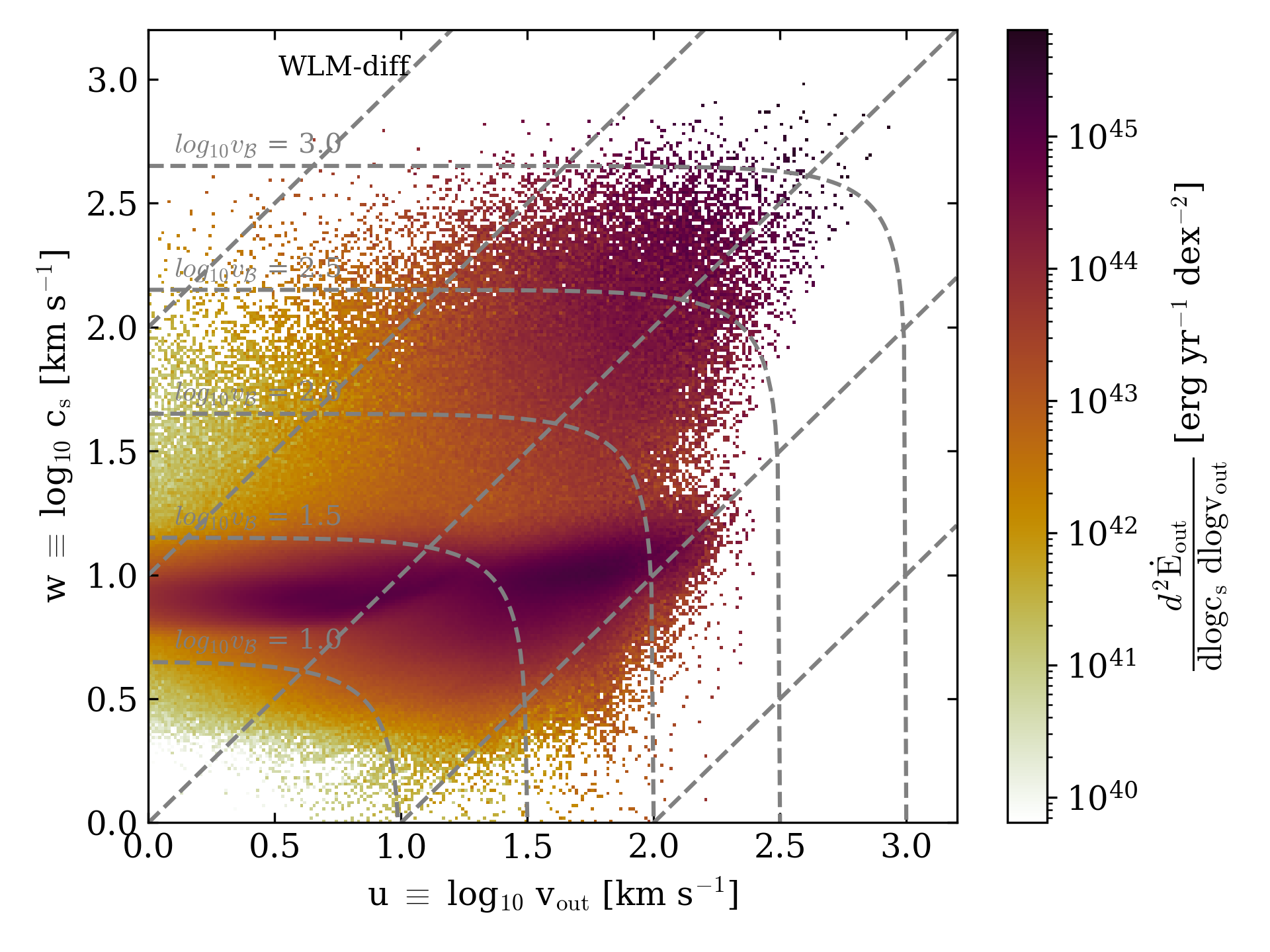}
    \includegraphics[width=0.49\textwidth]{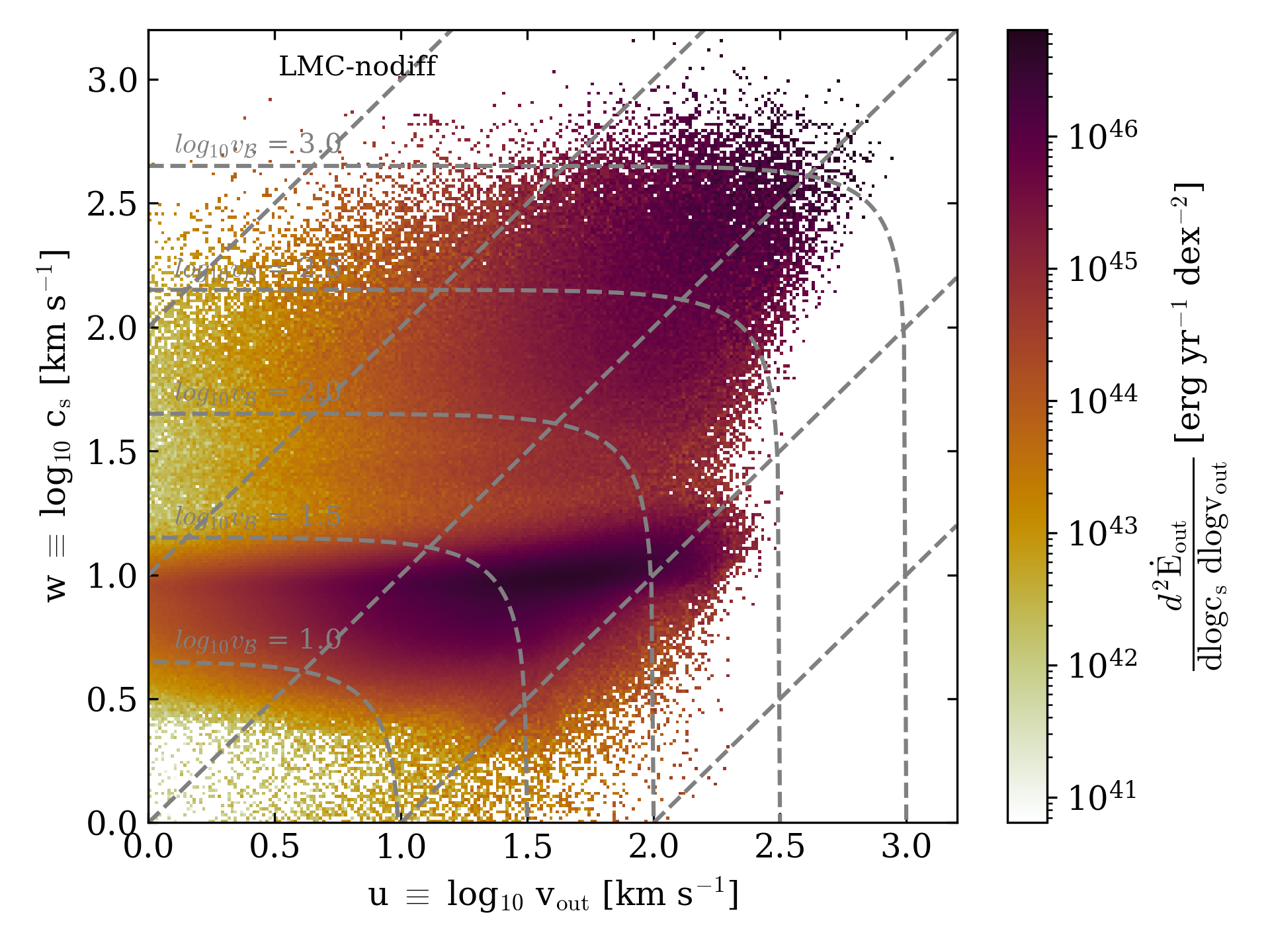}
    \includegraphics[width=0.49\textwidth]{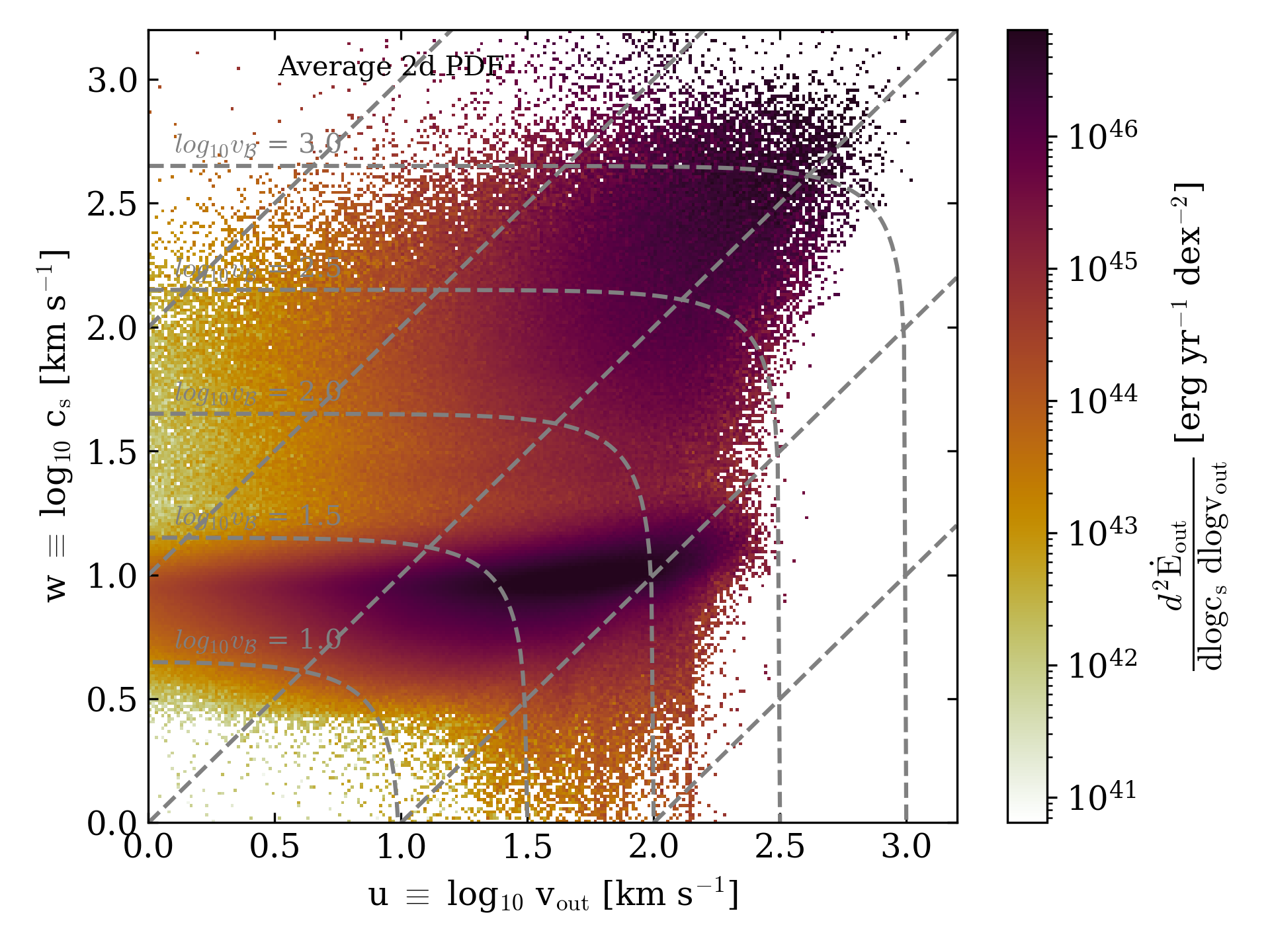}
    \caption{Joint 2D PDFs of outflow velocity and sound speed, color coded by the energy outflow rate for the models WLM-nodiff (top left), WLM-diff (top right), LMC-nodiff (bottom left) and LMC-diff (bottom right). We find that the sound speed in the hot phase of the wind is higher by a factor of two when metal diffusion is enabled.}
    \label{fig:energy_out}
\end{figure*}

\appendix

\section{Impact on the ISM phase structure}

In Fig.~\ref{fig:phase_temp}, we show classic density-temperature phase-space diagrams for the runs WLM-nodiff (left panel) and WLM-diff (right panel), time averaged over 400 Myrs (between 400 Myrs and 800 Myrs of temporal evolution of the system). We find that there is only a marginal impact on the shape of these. Visually, it appears that the ISM phase structure is ``smoother'' in the runs with metal diffusion, but the overall effects appear to be marginal. A similar statement can be made concerning the ISM's hot phase, which appears to be more mass-loaded in the metal diffusion runs. The clear bifurcation that can be seen in the metal mass in Fig.~\ref{fig:phase} is not present in these phase diagrams. The reason for this is that metals are subdominant in the total galactic mass budget, an effect that is simply averaged over in total mass weighted phase space diagrams presented in Fig.~\ref{fig:phase_temp}. We show this as  further evidence that the metal diffusion model only matters for very discrete effects that are related to the metal diffusion in the galaxy, an aspect of this work that is of key importance for studying metal abundance patterns in the ISM or the multiphase galactic outflow. 

\section{Impact on the mass- and energy-weighted phase structure of the outflow}

In Fig.~\ref{fig:mass_out}, we show the joint 2D PDFs of outflow velocity and sound speed, color coded by the mass outflow rate for the models WLM-nodiff (top left), WLM-diff (top right), LMC-nodiff (bottom left) and LMC-diff (bottom right). The impact of the metal diffusion model on the qualitative phase structure weighted by mass is negligible. However, we do note that the hot wind appears to carry slightly more mass in the models WLM-diff and LMC-diff, which is roughly a 10 percent effect for that phase. The reason why this does not appear in the integrated mass loading factor is that the star formation rate is about a factor of  two higher in the WLM-diff and LMC-diff models than in the WLM-nodiff and LMC-nodiff runs. It is interesting to note, however, that the sound speed in WLM-diff and LMC-diff is about a factor of two higher than in the corresponding models without metal diffusion.

In Fig.~\ref{fig:energy_out}, we show the joint 2D PDFs of outflow velocity and sound speed, color coded by the energy outflow rate for the models WLM-nodiff (top left), WLM-diff (top right), LMC-nodiff (bottom left) and LMC-diff (bottom right). The same trends occur as with the mass-weighted joint 2D PDFs, but the qualitative change of those and the impact of the metal diffusion model on the outflow energetics is negligible. 

Finally, we point out that the sound speed in the runs with metal diffusion WLM-diff and LMC-diff appears to be higher by around a factor of two than in the runs without metal diffusion, WLM-nodiff and LMC-nodiff.

\end{document}